\documentclass{ieeeaccess}
\usepackage{cite}
\usepackage{amsmath,amssymb,amsfonts}
\usepackage{algorithmic}
\usepackage{graphicx}
\usepackage{textcomp}

\usepackage{color, colortbl}
\definecolor{shadecolor}{gray}{0.9}
\usepackage{booktabs}
\usepackage{tabularx}
\usepackage{threeparttable}
\usepackage{comment}
\usepackage{multirow}
\usepackage[acronym]{glossaries}

\glsdisablehyper
\newacronym{ai}{AI}{Artificial Intelligence}
\newacronym{if}{IF}{Integrate and Fire}
\newacronym{iot}{IoT}{Internet of Things}
\newacronym{ann}{ANN}{Artificial Neural Network}
\newacronym{snn}{SNN}{Spiking Neural Network}
\newacronym{rsnn}{RSNN}{Recurrent Spiking Neural Network}
\newacronym{dnn}{DNN}{Deep Neural Network}
\newacronym{cnn}{CNN}{Convolutional Neural Network}
\newacronym{lif}{LIF}{Leaky Integrate and Fire}
\newacronym{asic}{ASIC}{Application-Specific Integrated Circuit}
\newacronym{fpga}{FPGA}{Field Programmable Gate Array}
\newacronym{stdp}{STDP}{Spike-Timing-Dependent Plasticity}
\newacronym{wta}{WTA}{Winner Takes All}
\newacronym{rtl}{RTL}{Register Transfer Level}
\newacronym{hdl}{HDL}{Hardware Description Language}
\newacronym{vhdl}{VHDL}{VHSIC Hardware Description Language}
\newacronym{sota}{SOTA}{State Of The Art}
\newacronym{shd}{SHD}{Spiking Heildelberg Digits}
\newacronym{sbs}{SbS}{Spike-by-Spike}
\newacronym{ml}{ML}{Machine Learning}
\newacronym{dl}{DL}{Deep Learning}
\newacronym{scnn}{SCNN}{Spiking Convolutional Neural Networks}
\newacronym{fc}{FC}{Fully-Connected}
\newacronym{fffc}{FF-FC}{Feed Forward Fully-Connected}
\newacronym{fsm}{FSM}{Finite State Machine}
\newacronym{MAC}{MAC}{Multiply and Accumulate}
\newacronym{BRAM}{BRAM}{Block RAM}
\newacronym{SRAM}{SRAM}{Static Random Access Memory}
\newacronym{DRAM}{DRAM}{Dynamic Random Access Memory}
\newacronym{LUT}{LUT}{Look Up Table}
\newacronym{FF}{FF}{Flip Flop}
\newacronym{DVS}{DVS}{Dynamic Vision Sensor}
\newacronym{SHD}{SHD}{Spiking Heidelberg Dataset}
\newacronym{BPTT}{BPTT}{Back-Propagation Through Time}
\newacronym{CU}{CU}{Control Unit}
\newacronym{ROM}{ROM}{Read Only Memory}
\newacronym{RAM}{RAM}{Random Access Memory}
\newacronym{DP}{DP}{Data Path}
\newacronym{CPS}{CPS}{Cyber-Physical System}
\newacronym{MLP}{MLP}{Multilayer Perceptron}
\newacronym{PE}{PE}{Portable Executable}
\newacronym{OS}{OS}{Operating System}
\newacronym{SBO}{SBO}{Stack Buffer Overflow}
\newacronym{ROP}{ROP}{Return-Oriented Programming}
\newacronym{JOP}{JOP}{Jump-Oriented Programming}
\newacronym{TP}{TP}{True Positive}
\newacronym{TN}{TN}{True Negative}
\newacronym{FP}{FP}{False Positive}
\newacronym{FN}{FN}{False Negative}
\newacronym{TPR}{TPR}{True Positive Rate}
\newacronym{FPR}{FPR}{False Positive Rate}
\newacronym{TNR}{TNR}{True Negative Rate}
\newacronym{ROC}{ROC}{Receiver Operating Characteristic}
\newacronym{AUC}{AUC}{Area Under the Curve}
\newacronym{HMD}{HMD}{Hardware-Supported Malware Detection}
\newacronym{HPC}{HPC}{Hardware Performance Counter}
\newacronym{PMU}{PMU}{Performance Monitoring Unit}
\newacronym{FE}{FE}{Feature Extraction}
\newacronym{FS}{FS}{Feature Selection}
\newacronym{PCA}{PCA}{Principal Component Analysis}
\newacronym{KNN}{KNN}{K-Nearest Neighbors}
\newacronym{SVM}{SVM}{Support Vector Machines}
\newacronym{HLS}{HLS}{High-Level Synthesis}
\newacronym{RADICS}{RADICS}{Rapid Attack Detection, Isolation and Characterization Systems}
\newacronym{TDT}{TDT}{Threat Detection Technology}
\newacronym{FCNN}{FCNN}{Fully Convolutional Neural Network}
\newacronym{FFT}{FFT}{Fast Fourier Transform}

\def\BibTeX{{\rm B\kern-.05em{\sc i\kern-.025em b}\kern-.08em
    T\kern-.1667em\lower.7ex\hbox{E}\kern-.125emX}}
\begin{document}
\history{\textcolor{white}{Date of publication xxxx 00, 0000, date of current version xxxx 00, 0000.}}
\doi{\textcolor{white}{XX.XXX/XXXX.XX.XX}}

\title{A survey on hardware-based malware detection approaches}

\author{
\uppercase{Cristiano Pegoraro Chenet}\authorrefmark{1}, \IEEEmembership{Student member, IEEE},
\uppercase{Alessandro Savino}\authorrefmark{1}, \IEEEmembership{Senior member, IEEE}, and
\uppercase {Stefano di Carlo}\authorrefmark{1}, \IEEEmembership{Senior Member, IEEE}.}
\address[1]{Department of Control and Computer Engineering, Politecnico di Torino, Corso Duca Degli Abruzzi, 24, 10129, Torino (TO), Italy (e-mails: {cristiano.chenet, alessandro.savino, stefano.dicarlo}@polito.it)}

\tfootnote{This work was partially supported by project SERICS (PE00000014) under the MUR National Recovery and Resilience Plan funded by the European Union - NextGenerationEU and by the Vitamin-V project (Project number: 101093062) funded by the European Union. Views and opinions expressed are, however, those of the author(s) only and do not necessarily reflect those of the European Union or the HaDEA. Neither the European Union nor the granting authority can be held responsible for them.}

\markboth
{Chenet et. al \headeretal: A survey on hardware-based malware detection approach}
{Chenet et. al \headeretal: A survey on hardware-based malware detection approach}

\corresp{Corresponding author: Stefano Di Carlo (e-mail: stefano.dicarlo@polito.it).}

\begin{abstract} 
This paper delves into the dynamic landscape of computer security, where malware poses a paramount threat. Our focus is a riveting exploration of the recent and promising hardware-based malware detection approaches. Leveraging hardware performance counters and machine learning prowess, hardware-based malware detection approaches bring forth compelling advantages such as real-time detection, resilience to code variations, minimal performance overhead, protection disablement fortitude, and cost-effectiveness.
Navigating through a generic hardware-based detection framework, we meticulously analyze the approach, unraveling the most common methods, algorithms, tools, and datasets that shape its contours. This survey is not only a resource for seasoned experts but also an inviting starting point for those venturing into the field of malware detection.
However, challenges emerge in detecting malware based on hardware events. We struggle with the imperative of accuracy improvements and strategies to address the remaining classification errors. The discussion extends to crafting mixed hardware and software approaches for collaborative efficacy, essential enhancements in hardware monitoring units, and a better understanding of the correlation between hardware events and malware applications.
\end{abstract}


\titlepgskip=-15pt

\maketitle

\section{Introduction}
\label{sec:introduction}

\PARstart{M}{alware}, short for malicious software, poses a significant threat to computer security. It includes any code modification within a software system aimed at causing harm or disrupting the system's intended function \cite{McGraw_Morrisett_2000, Alsmadi:2021aa}. Malware attacks cover spying, intrusive ads, email abuse, system damage, ransom demands, data release, slowdown, browser manipulation, and unauthorized access to sensitive information. Successful attacks lead to consequences that can be categorized into four groups: (i) unauthorized disclosure, where an authorized entity gains access to data; (ii) deception, where an authorized entity receives false data; (iii) disruption, causing interruptions in system services; and (iv) usurpation, resulting in unauthorized control of system services \cite{Stallings_Brown_2014}.
Computing systems, including personal computers, mobile phones, \gls{iot}, 5G devices, \glspl{CPS}, and enterprise-wide systems, are vulnerable to malware. The complexity and size of modern systems, often indicated by a rising number of lines of code, amplify the threat. Factors such as numerous bugs, unsafe programming languages, improper configuration, and the ease of concealing malicious code create potential vulnerabilities. Additionally, the increased network connectivity expands the security risks, making all devices potential targets for attackers. For example, cybercrimes have seen a 70\% increase in online fraud accomplished through mobile platforms, with a 30\% rise in \gls{iot} malware in 2020 \cite{SonicWall_2020}.

Globally, cybersecurity is paramount, with malware being a primary vehicle for cybercrimes. The World Economic Forum Global Risk Report 2023 ranks cyber insecurity eighth among top global risks, alongside threats like climate change and involuntary migration \cite{GlobalRisckReport2023}. Cybersecurity Ventures predicts a 15 percent annual growth in international cybercrime costs, reaching USD 8 trillion in 2023 and USD 10.5 trillion annually by 2025 \cite{GlobalRisckReport2023}. Global spending on cybersecurity products and services is expected to exceed USD 1.75 trillion from 2021 to 2025, growing 15 percent year-over-year \cite{cybersecurity_almanac_2023}. Ransomware, a prevalent malware threat, was predicted to cost USD 20 billion globally in 2021, with damage costs projected to exceed USD 265 billion annually by 2031 \cite{GlobalRisckReport2023}. 

Researchers have developed various malware detection methods in response to these alarming statistics, leveraging \gls{ml} and \gls{dl} techniques. Surveys have evaluated and categorized research in this domain, focusing on specific \glspl{OS}, such as Windows, or mobile platforms like Android.
Ye et al. \cite{Ye:2017aa} conducted a comprehensive survey on intelligent malware detection using data mining techniques, emphasizing the importance of \gls{FE} and algorithm selection. Subsequently, Ucci et al. \cite{Ucci:2019aa} provided an overview of machine learning-based malware analysis, focusing on analysis objectives, \gls{FE}, and \gls{ml} algorithms, albeit limited to \gls{PE} files.
Gibert et al. \cite{Gibert:2020aa} systematically reviewed \gls{ml} and \gls{dl} techniques for Windows malware detection, comparing input features, classification algorithms, and dataset characteristics. Similarly, Qiu et al. \cite{Qiu:2020aa} and Liu et al. \cite{Liu:2022aa} addressed deep Android malware detection, emphasizing supervised classification using \glspl{MLP} and \glspl{cnn} architectures.
Catal et al. \cite{Catal:2022aa} conducted an extensive literature review on \gls{dl} techniques for mobile malware detection, highlighting the prevalence of \gls{MLP} and \gls{cnn} architectures, with a focus on supervised learning and static features.
Furthermore, Deldar et al. \cite{Deldar:2023aa} proposed a survey on \gls{dl} techniques for zero-day malware detection, targeting features extracted at the software level to address emerging threats.

In the early 2010s, researchers initially proposed the idea of \gls{HMD} \cite{Malone_2011, Demme_2013}. \gls{HMD} involves dynamically analyzing micro-architecture events in a processor using \gls{ml} algorithms to differentiate between benign applications and malware. The shift towards \gls{HMD} is justified because of the potential of enhanced security by leveraging robust hardware monitoring infrastructures. This provides a more robust defense against sophisticated attacks that may exploit vulnerabilities in software-based approaches. Specifically, hardware features reflect phase behavior in the underlying hardware, as observed in prior studies \cite{Sherwood_2003, Isci_2006}. These phases often correspond to time-behavioral patterns in micro-architectural events, which vary significantly between programs, enabling the distinction between malicious and benign applications. Additionally, these hardware-based approaches address the zero-day issue, as demonstrated in \cite{He:2021aa}. To the best of our knowledge, a comprehensive overview of \gls{HMD} methods is still missing. This paper tries to cover this gap.

The structure is as follows: Section \ref{sec:malware_basics} covers the basics of malware, serving as a foundation for understanding the field. Section \ref{sec:malware_detection} presents a comprehensive overview of software and hardware-based malware detection solutions, with a detailed discussion of their strengths and weaknesses. Section \ref{sec:hw_based_detection_app} delves into crucial aspects of hardware-based detection. Lastly, Section \ref{sec:conclusion_challenges} provides conclusions and outlines research challenges. 
\section{Malware Fundamentals}
\label{sec:malware_basics}

Categorizing malware is difficult because of its growing complexity and diverse properties. Yet, creating a malware taxonomy provides valuable insights into understanding it better. Before exploring the fundamentals of malware operation, let us define a set of keywords commonly used to describe different malware categories \cite{McGraw_Morrisett_2000,Christodorescu_2007}:

\begin{itemize}
	\item \textbf{Virus}: malicious code with the capability of inserting itself into other programs;
 	\item \textbf{Worm}: malicious code that propagates similarly to viruses but does not require a target software to replicate, often exploiting connectivity such as emails;
	\item \textbf{Trojan horse}: malicious code that masquerades as a useful program;
	\item \textbf{Spyware}: malicious code secretly installed into an information system to transmit private user data to an external entity;
	\item \textbf{Adware}: malicious code that displays computer advertisements, primarily aiming for financial benefits;
	\item \textbf{Ransomware}: malicious code that denies access to a user’s data, usually by encrypting it until a ransom is paid;
	\item \textbf{Backdoor}: malicious code that opens systems to external entities by subverting local security policies to allow remote access and control over a network;
	\item \textbf{Keylogger}: malicious code designed to record keystrokes, used to obtain passwords or encryption keys to bypass security measures;
	\item \textbf{Botnet}: a network of infected computers controlled by a remote criminal;
	\item \textbf{Rootkit}: malicious application attackers use to conceal their activities and maintain control over a host.
\end{itemize}

Organizations like NIST \cite{NIST_Glossary_2023} and ENISA \cite{ENISA_Botnets_2023} recognize these malware types. In literature, three common properties describe malware: (i) propagation method, categorizing based on spread and purpose; (ii) concealment strategy, focusing on hiding tactics against users and detection; and (iii) data structure manipulation, dealing with software vulnerability exploitation. Table \ref{tab:classifiers_analysis} organizes malware based on these categories.

\begin{table*}[hbt]
 \centering
\caption{Malware categories based on propagation method, concealment strategy, and data structure manipulation.}\label{tab:malware_classification}
\begin{tabular}{lccc}
\toprule
\textbf{Malware Type} & \textbf{Propagation Method} & \textbf{Concealment Strategy} & \textbf{Data Structure Manipulation} \\
\midrule\\
\rowcolor{shadecolor}
Worms & Network-based transmission & Polymorphisms or metamorphism & Exploitation of memory corruption vulnerabilities \\
\midrule
Viruses & File-based transmission & Polymorphism & Manipulation of data structures \\
\midrule
\rowcolor{shadecolor}
Trojans & Social engineering & No concealment & - \\
\midrule
Spyware & Internet downloads & Encryption & - \\
\midrule
\rowcolor{shadecolor}
Ransomware & Email attachments & Encryption & File system manipulation \\
\midrule
Adware & Software bundling & No concealment & - \\
\midrule
\rowcolor{shadecolor}
Rootkits & Kernel-level exploits & Obfuscation & Manipulation of system structures \\
\midrule
Backdoors & Remote access & Encryption & - \\
\midrule
\rowcolor{shadecolor}
Keyloggers & Phishing, infected software & Encryption & - \\
\midrule
Botnets & Exploitation, social engineering & Encryption and polymorphism & - \\
\bottomrule
\end{tabular}
\end{table*}

Regarding concealment strategy, malware can be categorized into two main groups: (i) no concealment and (ii) stealthy malware \cite{Aycock_2010,You_Yim_2010,Rad_2012}. No concealed malicious code lacks techniques to hide itself, making it easy to detect. However, as shown in Table \ref{tab:malware_classification}, only a small subset of malware does not employ concealment. File infectors like traditional viruses or worms may not heavily focus on concealment, spreading by attaching to executable files. Adware may not invest heavily in hiding and may rely on user interactions. Similarly, if achieved without sophisticated evasion, simple trojans may prioritize their primary goal over concealment.

Conversely, stealthy malware is a general term for all kinds of malicious code capable of hiding from users and detection mechanisms \cite{Stolfo_2007, Rudd_2017}. Its primary purpose is to remain undetected for an extended period in the computing system, allowing compromising computers and stealing information before a suitable detection mechanism can be deployed to protect against it. In general, the concealment actions aim to hide the malware's trails or code. Stealthy malware may employ several techniques:

\begin{itemize}
    \item \textbf{Encryption/obfuscation}: the oldest and simplest technique consists of a decryptor and an encrypted main body. When the infected file runs, the decryptor recovers the main body. The malware may use a different key for each infection to hide its signature, making the encrypted part unique. The decryptor small size compared to the main body reduces detection probability. Encryption complexity ranges from basic operations to strong encryption methods \cite{Aycock_2010,You_Yim_2010, Nadim:2021aa};
    
    \item \textbf{Oligomorphism and polymorphism}: the encryption technique limitation lies in the constant decryptor across exploitations, enabling detection based on code patterns. Oligomorphism employs a small set of decryptors, using a different one for each infection. Polymorphism, similar but with theoretically infinite decryptor variations, relies on obfuscation methods like dead-code insertion and register reassignment for distinct decryptor creation \cite{Aycock_2010,You_Yim_2010,Wong_Stamp_2006,Konstantinou_2008};
    
    \item \textbf{Metamorphism}: the binary sequence is altered by making a new malware version for each new infection through a mutation engine. The mutation engine uses code transforming and obfuscation to change the malicious code \cite{Aycock_2010,You_Yim_2010,Brezinski_2023}.
 \end{itemize}

Several classes of software vulnerabilities can be explored to perform security attacks. This paper focuses on the prevalent memory errors enabling memory corruption for security attacks \cite{Van_der_Veen_2012}, which lead to two main exploit categories: control-flow attacks and data-only attacks.

Control-flow attacks are common, easy to construct, and demand minimal application-specific knowledge. They exploit vulnerabilities like buffer overflows or injection attacks to redirect the program's execution flow, enabling arbitrary code execution \cite{Demme_2013, Khasawneh_2015, Ozsoy_2015, Tang_2014, Wang_Karri_2013}. Techniques such as code injection \cite{Ray_Ligatti_2012}, \gls{ROP} \cite{Prandini_Ramilli_2012}, or \gls{JOP} \cite{Bletsch_2011} divert execution to specific memory locations housing malicious code, bypassing standard security measures.

In contrast, data-only attacks are rarer, subtler, and require advanced knowledge of program semantics. They manipulate critical data while maintaining a valid control flow, compromising target programs without injecting additional code. These attacks alter essential data elements, such as identification or configuration data, influencing target application behaviors during runtime \cite{Chen_2005}.
\section{Overview of Malware detection}
\label{sec:malware_detection}

Malware detection involves determining whether a given program exhibits malicious intent. Figure \ref{fig:overview_malware_detection_approaches} offers an overview of contemporary solutions for malware detection, categorized into two main groups: software-based and hardware-based approaches. This division is rooted in differing observation points within the system stack and different detection methodologies. Recent advancements, as underscored by \cite{Deldar:2023aa} and \cite{He:2021aa}, increasingly rely on \gls{ml} or \gls{ai} techniques to facilitate detection.


\Figure[htb]()[width=0.98\columnwidth]{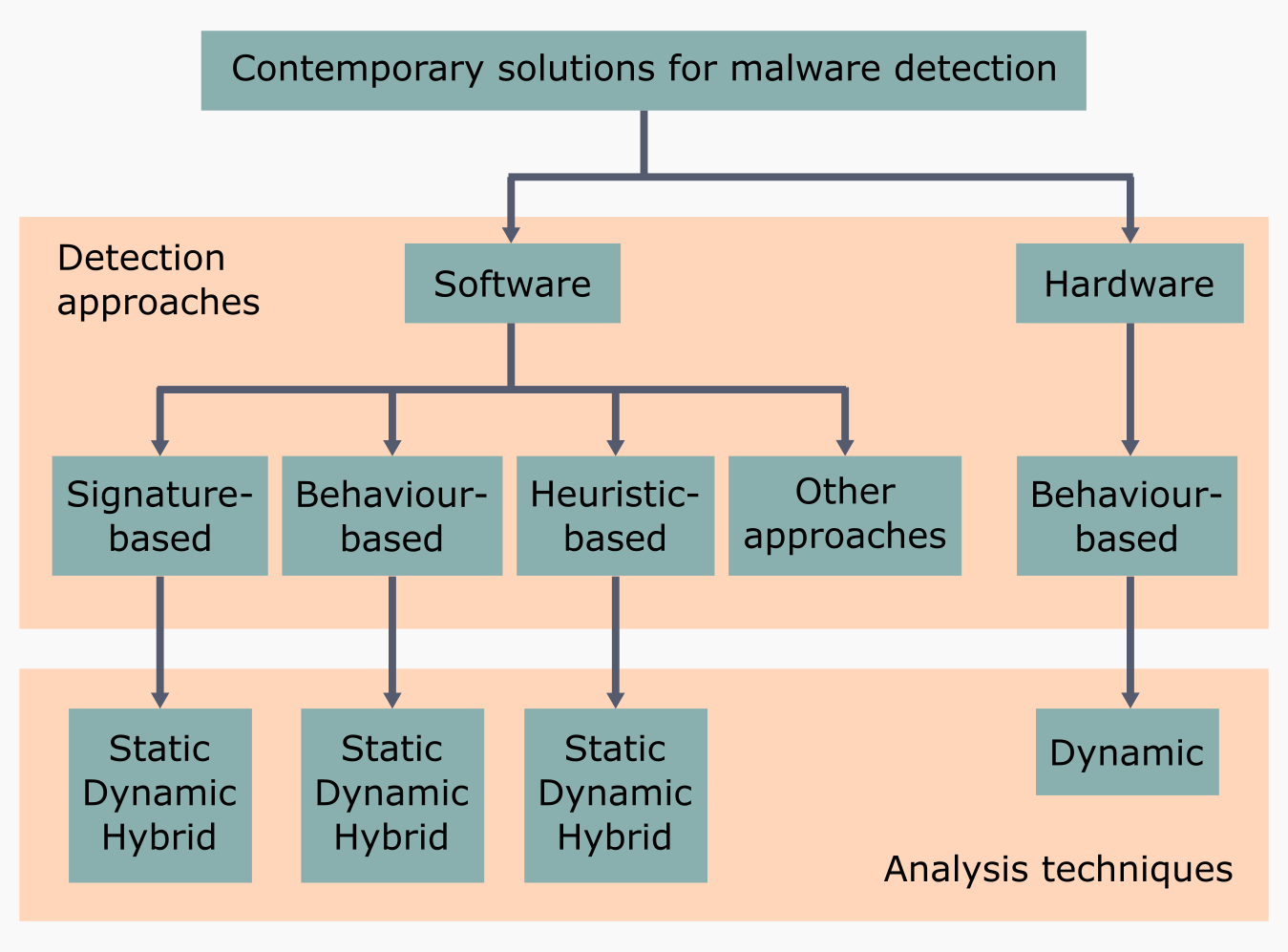}
   {Overview of the contemporary solutions for malware detection. Elaborated by the authors based on \cite{Aslan_Samet_2020}.\label{fig:overview_malware_detection_approaches}}

This section presents an overview of software-based and hardware-based malware detection (sub-sections \ref{subsec:software_malware_detection} and \ref{subsec:hardware_malware_detection}), starting by reviewing the metrics used for evaluating the performance and efficiency of the detectors (sub-section \ref{subsec:evaluation_metrics}). 

\subsection{Evaluation metrics}
\label{subsec:evaluation_metrics}

Before delving into specific malware detection techniques, readers need to consider the evaluation metrics used to assess their effectiveness. These metrics serve as quality indicators, pivotal in determining the adoption of a technique on a commercial scale. Since malware detection is a classification problem, the quality evaluation of the detectors is based on the standard classification metrics. They can be grouped as performance metrics and efficiency metrics. Performance is the degree to which a system or component accomplishes its designated functions within given constraints, i.e., correctly detects the malware. Efficiency is the degree to which a system or component performs its specified functions with minimum consumption of resources \cite{ISO_IEC_IEEE_Vocabulary_2017}. 

The primary evaluation tool for performance is the confusion matrix. This matrix is fundamental in \gls{ml} and classification tasks, summarizing results in a tabular form. It comprises four elements (see Table \ref{tab:confusion_matrix}): \glspl{TP} represent instances where the model correctly predicts malware presence, \glspl{TN} indicate correct predictions of malware absence. In contrast, \glspl{FP} and \glspl{FN} denote incorrect predictions of malware presence or absence, respectively.

\begin{table}[htb]
\centering
\caption{Confusion matrix for malware detection.}
\label{tab:confusion_matrix}
\begin{tabular}{ccc}
\toprule
 & \textbf{Predicted Negative} & \textbf{Predicted Positive} \\
\midrule\\
\rowcolor{shadecolor}
\textbf{Actual Negative} & \glspl{TN} & \glspl{FP} \\
\midrule
\textbf{Actual Positive} & \glspl{FN} & \glspl{TP} \\
\bottomrule
\end{tabular}
\end{table}

Such a matrix allows for the definition of more descriptive metrics, and Table \ref{tab:usefull_matrix} summarizes the most common ones \cite{Ye:2017aa}. The \textit{accuracy} summarizes the overall correctness of the classification model by expressing the number of correct predictions, making it one of the most widely used metrics. In scenarios where it is crucial to avoid incorrect malware predictions, \textit{precision} provides an accurate measure of the \glspl{TP} among all positive predictions. Shifting the evaluation focus to ensure no malware passes unnoticed, the \textit{\gls{TPR}} (also known as \textit{Recall} or \textit{Sensitivity}) weighs \glspl{TP} against all positive samples. It has two counterparts: (i) the \textit{\gls{FPR}}, representing the probability of a \gls{TP} being missed, and (ii) the \textit{specificity}, also known as \textit{\gls{TNR}}, indicating the probability of an actual negative (\gls{TN}) being correctly classified. Balancing Precision and Recall is often essential, and the evaluation can be accomplished using the \textit{F1-score}, which represents their harmonic mean.

Eventually, the \textit{\gls{ROC} curve} offers a visual perspective to performance evaluation. It plots the \gls{TPR} against the \gls{FPR} on a 2D graph, enabling a visual comparison of different models and capturing multiple classification aspects by inspecting the \textit{\gls{AUC}}. In simple terms, the larger the \gls{AUC}, the better the model. \gls{AUC} is closely related to the robustness of the classifier, indicating how effectively the classifier distinguishes between malware and benign applications.

\begin{table}[htb]
\centering
\caption{Most common metrics for performance evaluation of classification.}
\label{tab:usefull_matrix}
\begin{tabular}{cc}
\toprule
 \textbf{Matrix} & \textbf{Expression} \\
\midrule\\
\rowcolor{shadecolor}
\textbf{Accuracy (A)} & $A=\frac{TP+TN}{TP+TN+FP+FN}$ \\
\midrule
\textbf{Precision (P)} & $P=\frac{TP}{TP+FP}$ \\
\midrule
\rowcolor{shadecolor}
\textbf{\gls{TPR}} & $TPR=\frac{TP}{TP+FN}$ \\
\midrule
\textbf{\gls{FPR}} & $FPR=\frac{FN}{FN+TP}$ \\
\midrule
\rowcolor{shadecolor}
\textbf{Specificity (S)} & $S = \frac{TN}{TN+FP}$ \\
\midrule
\textbf{F1-Score (F1)} & $F1 = \frac{2 \times (P \times R)}{(P + R)}$ \\
\midrule
\rowcolor{shadecolor}
\textbf{ROC} & $ROC = 1 - S = \frac{FP}{FP+TN}$\\
\midrule
\textbf{AUC} & $AUC = \int_{0}^{1} R (FPR) \,dFPR$\\
\bottomrule
\end{tabular}
\end{table}

According to \cite{ISO_IEC_IEEE_Vocabulary_2017}, efficiency is related to the resources used for malware detection. Many metrics can be used to evaluate the efficiency \cite{Sze_2020}, but in the malware detection field, latency, power consumption, and hardware cost are the main interest:

    \begin{itemize}
    
    \item \textbf{Latency} is the time between collecting all features analyzed by the malware detector and concluding its detection. A low latency is vital for run-time detection of malware that acts in a short interval of time;
   
    \item \textbf{Power consumption} indicates the energy the detector consumes per unit of time. Two factors primarily impact the power consumption of the detector: the hardware that implements or where the classifier runs and the detection algorithm (those with higher computing processing tend to consume more);

    \item \textbf{Hardware cost} indicates the monetary cost of building the detection system. This is important from both an industry and a research perspective to dictate whether a system is ﬁnancially viable. The main parameter to evaluate the hardware cost is the chip area (usually reported in square millimeters) in conjunction with the process technology (for example, 45 nm). Sometimes, the amount of memory is also used to evaluate the hardware cost. 
    \end{itemize}



\subsection{Software-based malware detection}
\label{subsec:software_malware_detection}

Software-based protection relies on specific software running in the system and analyzing the potential malware presence using different approaches. Authors in \cite{Aslan_Samet_2020} and \cite{Deldar:2023aa} proposed a very comprehensive selection of them:

\begin{itemize}

    \item \textbf{Signature-based}: the signature is a unique malware feature extracted from structural properties (e.g., code sequences) or run-time properties \cite{Idika_Mathur_2007}. The detection works as follows: features extracted from the executable generate a signature stored in a signature database. When the system is required to classify a potential threat, the detector extracts the related features and computes the signature, comparing it with signatures on the database. The potential threat is marked as malware if a hit occurs during the comparison. This approach is widely used within commercial antivirus and does not allow zero-day detection~\cite{Deldar:2023aa};

    \item \textbf{Software behavior analysis}: this approach is based on dynamic characteristics from run-time executions of programs \cite{Aslan_Samet_2020}. Dynamic characteristics might include processor and memory information, kernel usage (system calls), file system activities, and network communications. They are extracted with monitoring tools, a dataset is created, and a \gls{ml} detector distinguishes malicious and harmless applications. Software behavior analysis can detect malware variants often missed by the signature-based approach;


    \item \textbf{Heuristic-based detection}: this method relies on experiences and techniques, including rules and \gls{ml}. The process involves two phases: first, the detector system is trained with normal and abnormal data to identify relevant characteristics. In the second phase, known as monitoring or detection, the trained detector intelligently assesses new samples to make decisions \cite{Alzarooni_2012};
    
    \item \textbf{Deep Learning}: this falls under the umbrella of \gls{ml} algorithms, enabling computational models with multiple layers to extract more advanced features from raw input~\cite{Deldar:2023aa}. The \gls{FE} aspect combines elements from previous approaches, making it a novel method. Additionally, it proves highly effective for zero-day detection, as the \gls{FE}, employing multiple techniques, facilitates context adaptation and model updates, as highlighted in \cite{Deldar:2023aa}.
    \end{itemize}

Regarding software-based detection, it is also crucial to distinguish among the types of analysis carried out to extract the required information. According to \cite{Idika_Mathur_2007}, three ways are possible: (i) via \textit{static} analysis, using syntax or structural properties of the program/process (e.g., code sequences), (ii) via \textit{dynamic} analysis, extracting the necessary data during or after program execution, leveraging run-time information, and (iii) via \textit{hybrid} analysis, combining the two previous. Selecting one of those also affects the expected latency of the detection. While a static analysis aims to detect the threat even before executing the malicious program, the other two might require an entire execution before detection.

\subsection{Hardware-based malware detection}
\label{subsec:hardware_malware_detection}

Hardware-based detection, or \gls{HMD}, addresses the performance and computational overhead challenges of traditional malware detection techniques by utilizing low-level micro-architectural features of running applications on the target system \cite{He:2021aa}. The concept that malware can be identified through micro-architecture hardware events stems from the observation that programs exhibit phase behaviors \cite{Sherwood_2003, Isci_2006}. Program phases, which vary significantly between programs, manifest as patterns in architectural and micro-architectural events. This variation enables the discrimination of programs based on their time-behavioral hardware event patterns, facilitating the differentiation between malicious and benign applications. In 2011, Malone et al. \cite{Malone_2011} demonstrated the feasibility of detecting program code modifications based on the deviation of hardware events. In 2013, Demme et al. \cite{Demme_2013} showed the feasibility of detecting Android malware and Linux rootkits using hardware events values analyzed by a \gls{ml} classifier. 

The idea of \gls{HMD} is to perform dynamic analysis leveraging micro-architecture hardware events monitored by most modern microprocessors using \glspl{HPC} \cite{Alonso:2023aa}. Various \gls{ml} techniques can be applied to the \glspl{HPC} collected data \cite{He:2021aa}. 
One of the primary advantages of \gls{HMD} is that the analysis relies on real-time hardware collected data, enabling fast \gls{ml} classification; a few milliseconds suffice to identify threats. This translates to low latency, enabling runtime detection \cite{Sayadi_2018, Sayadi_2019, Patel_2017}. Unlike static technique analysis employed by most software-based antivirus solutions, which can be easily subverted by stealthy malware using concealment techniques, dynamic analysis via hardware-based approaches facilitates the detection of code variants and unknown malware \cite{Demme_2013}. Moreover, while software-based detection tools are software-based and susceptible to bugs or oversights in the underlying system software, hardware-based detection with secure hardware significantly reduces the possibility of malware subverting protection mechanisms \cite{Demme_2013, Tang_2014}.

On the performance front, the dynamic analysis conducted by software-based detection necessitates sophisticated computation, often at the expense of significant performance overhead. The increasing software size further complicates dynamic software analysis \cite{Demme_2013}. Conversely, in the hardware-based approach, understanding software behavior provided by micro-architectural events simplifies the analysis, reducing computational processing efforts and the cost of hardware-based detection \cite{Demme_2013, Sayadi_2022}.


deHowever, while the \glspl{HPC} demonstrate their ability to track behavioral deviations~\cite{Dutto:2021aa, Torres_Liu_2022, Kasap:2023aa}, their effectiveness remains open to discussion. On the positive side, \cite{Demme_2013, Tang_2014} demonstrated detector performance using this approach, reporting accuracy consistently exceeding 80\%, deeming it effective. Conversely, \cite{Zhou_2018} and \cite{Zhou_2021} conducted experiments challenging the effectiveness of hardware-based detection. They argued that reported detection capabilities often stem from tiny sample sizes and experimental setups favoring the detection mechanism unrealistically. Even if accurate, an 80\% accuracy is insufficient in scenarios with thousands of executables, risking many benign applications being misclassified as malware. They also questioned the causal link between low-level micro-architectural events and high-level software behavior. Lastly, they illustrated the hardware-based detector inability to distinguish ransomware embedded in a benign application like Notepad++. In a recent contribution, \cite{Botacin_Gregio_2022} acknowledged the absence of a perfect malware detector and argued that hardware-based detection is only effective for specific malware types. In particular, \cite{Botacin_Gregio_2022} proposes its effectiveness in identifying attacks exploiting architectural side-effects, citing examples such as RowHammer \cite{Kim_2014, Mutlu_2020} (detectable through excessive cache flushes \cite{Li_Gaudiot_2019}), \gls{ROP} attacks \cite{Prandini_Ramilli_2012} (identified by an abundance of instruction misses \cite{Wang_Backer_2016}), and DirtyCoW \cite{NIST_CVE-2016-5195} (detectable through heightened paging activity). The authors also emphasized the necessity for a maliciousness theory to enhance the understanding of malware threats and assess proposed defenses.

While \glspl{HPC} have been used in the past for safety and security, performance analysis, and optimization \cite{Weaver_McKee_2008,Carelli:2018aa,Carelli:2019aa}, it is well-known that they may suffer from inconsistency in implementation, leading to non-determinism and overcounting \cite{Weaver_Terpstra_Moore_2013}. Das et al. highlighted some of these \gls{HPC} challenges in security \cite{Das_2019}. Recent studies address \gls{HPC} discrepancies, propose methodologies, analyze resilience, and compare HPCs in various machines \cite{Barrera_2020,Kadiyala_2020,Ritter_2022,Sasongko_2023}. Given that \glspl{HPC} are hardware-based protections, detectors may be designed for specific devices with characteristics defined by the architecture and manufacturer. For instance, processors may track different numbers of events simultaneously, and discrepancies in instruction counting methods are possible \cite{Weaver_McKee_2008}. These factors underscore the need for malware detection applications to abstract software from the hardware level.

Among the inconsistencies and limitations of \glspl{HPC}, some countermeasures can be deployed to stabilize the generated data~\cite{Weaver_McKee_2008, Das_2019}. They include per-process filtering of events (applied by saving and restoring the counter values at context switches), proper interrupt handling, and minimizing the impact of non-deterministic events. In general, all works acknowledge that the evolution and improvement of the processors hardware monitoring units also tend to reduce this issue. 
Eventually, the classification task built on top of the \gls{HPC} data is commonly a \gls{ml} one. This frequently leads to techniques that increase the complexity of such algorithms, like ensemble learning and time series or even \glspl{dnn}~\cite{He:2021aa}.

\section{Hardware-based Malware Detection Basics}
\label{sec:hw_based_detection_app}

This section focuses on \gls{HMD} techniques, outlining their key components. 

\subsection{Hardware events and performance counters}
\label{subsec:hw_events_perf_count}

Modern processors have units to monitor hardware events. In 2002, Sprunt \cite{Sprunt_2002} published a seminal paper on the basics of \glspl{PMU}. These units were developed to collect data about the performance of applications, operating systems, and processors and to help programmers tune algorithms and codes. Software dynamically adjusted to resource utilization would also benefit from the information collected. The proven advantages of utilizing the \glspl{PMU}, the continuous improvements of these units, and their constant spreading among different devices have led to their leverage for safety and security purposes~\cite{Dutto:2021aa,Kasap:2023aa,Carelli:2018aa,Carelli:2019aa}.

Nowadays, \glspl{PMU} can monitor several hardware events (see Figure \ref{fig:HW_events_counters}). Complex devices like high-end processors have hundreds of events to monitor. These events include retired instructions (branches, load, store, etc.), branch predictions, cache hits and misses, floating-point operations, hardware interrupts, elapsed core clock ticks, core frequency, and temperature. However, to minimize hardware complexity, only a few \glspl{HPC} (e.g., 2 to 8 in high-end processors) are generally available, thus limiting the number of parallel events that can be monitored. Each \gls{HPC} has an event detector and an associated counter \cite{Doyle_2017}.

\Figure[htb]()[width=0.98\columnwidth]{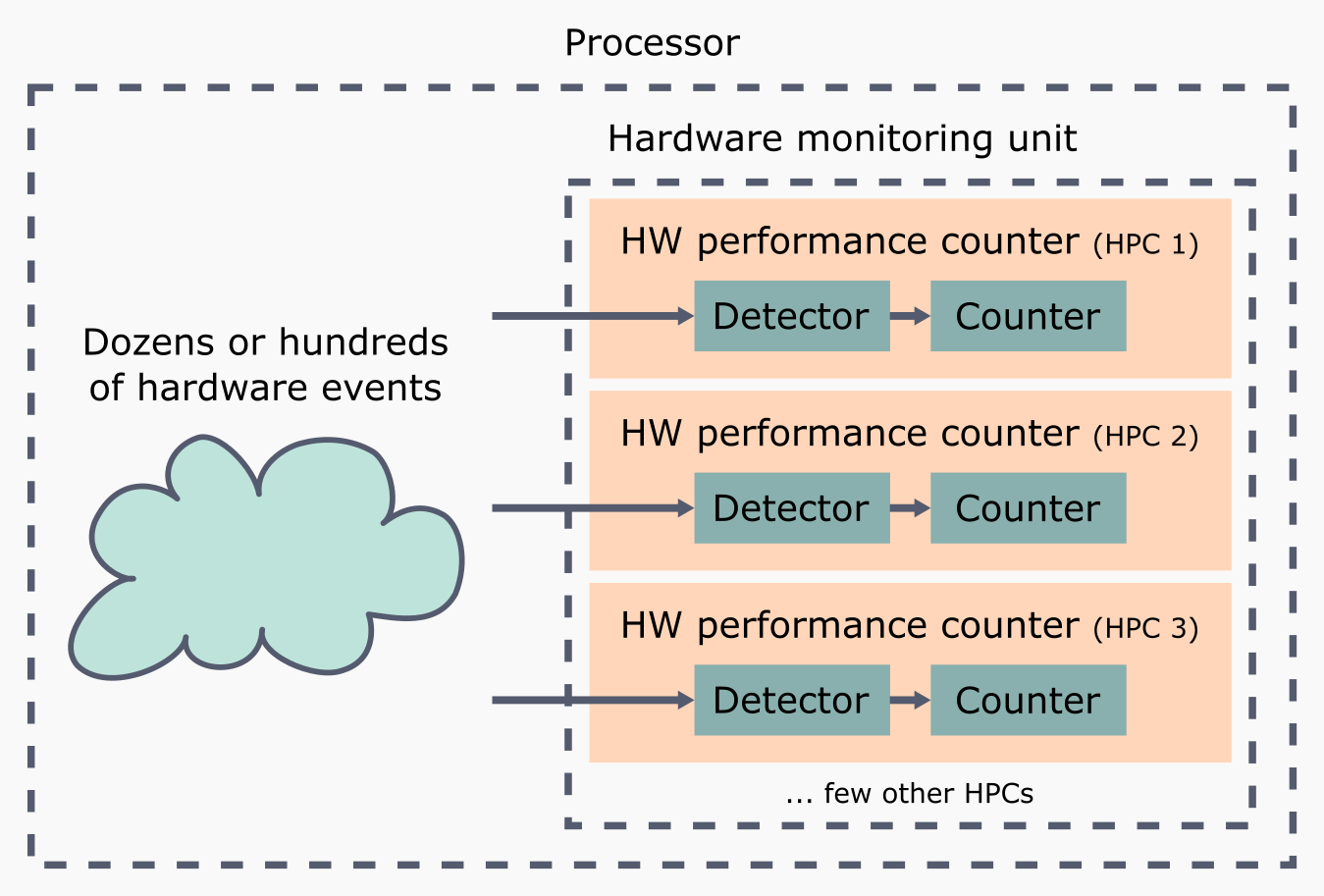}
   {Hardware events and performance counters in a processor. Elaborated by the author. \label{fig:HW_events_counters}}

\subsection{Hardware-based detection framework}
\label{subsec:hardware-based_detection_framework}

A generic framework can be a guiding structure to facilitate the implementation of \gls{HMD}, as illustrated in Figure \ref{fig:generic_hardware-based_framework}. The framework leverages the existing \gls{PMU} within the processor and consists of two primary components: (i) data collection and preprocessing and (ii) malware detection. This section provides a detailed overview of the implementation process.

\Figure[htb]{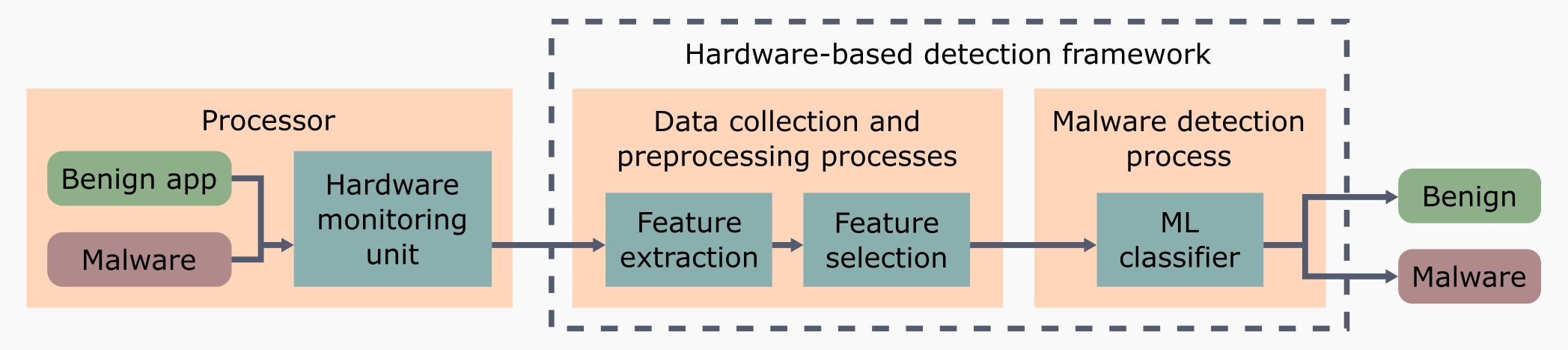}
   {A generic hardware-based detection framework. Elaborated by the author. \label{fig:generic_hardware-based_framework}}
   
Data collection involves \gls{FE} and \gls{FS} \cite{Abdulwahab_2022, Chandrashekar_Sahin_2014}. \gls{FE} captures and stores \glspl{HPC} in a vector space, enabling the \gls{FS} to select a subset that efficiently describes the input data while minimizing noise and irrelevant variables, ensuring optimal prediction results. \gls{FE} can occur in the time or event domain \cite{Sprunt_2002}. In the time-based domain, the application execution is periodically interrupted to record \gls{HPC} values. Conversely, the event domain triggers interruptions based on specific events or a set number of executed instructions rather than regular intervals.

In terms of strategies to perform \gls{FE}, we envision four alternatives: (i) instrument the source code with the employment of a library, like \texttt{PAPI} \cite{PAPI_1999}; (ii) develop of a proprietary kernel module or driver, as performed in \cite{Tang_2014}; (iii) use of an available utility that performs tasks mainly in the \gls{OS} kernel, like \texttt{PERF} \cite{PERF_2009}; and (iv) use of a micro-architectural simulator to model the processor as it executes the application, like \texttt{gem5} \cite{Binkert_2011} and \texttt{GVSoC} \cite{Bruschi_2021}.

During \gls{FE}, the sampling strategy is crucial. In the time-based domain, parameters such as period, frequency, or number of cycles determine when \glspl{HPC} are sampled. In the event-based domain, sampling depends on the number of event or instruction occurrences. The chosen \gls{FE} strategy influences these definitions. A proprietary kernel module or driver allows programmers to choose between time-based or event-based domains, set parameters for sampling triggering, and specify values. However, configurations are limited when libraries like \texttt{PAPI} and \texttt{PERF} are used.
Regarding sampling values, in time-based sampling, there is no fixed ideal period or frequency, varying based on the experiment and goal. Hardware-based detection experiments typically use periods in the order of milliseconds or seconds. Striking a balance between low and high sampling frequencies is essential, considering the trade-off between computational processing, data quantity, and system effects.

\gls{FS} offers multiple advantages, including addressing the Curse of Dimensionality in \gls{ml} \cite{Goodfellow_2016}, enhancing data understanding, reducing computation requirements, and improving predictor performance. Filter-based algorithms dominate the \gls{FS} in the \gls{HMD} field, ranking features based on a scoring criterion, using a threshold for variable selection. They are valued for simplicity and practical application success, focusing on the relevancy of features. Prominent methods include \gls{PCA} (used by \cite{Zhou_2018, Sayadi_2019, Gao_2021, Sayadi_2021}), Fisher Score \cite{Duda_2000} (used by \cite{Tang_2014, Torres_Liu_2022}), Pearson Correlation Coefficient \cite{Pearson_1895} (used by \cite{Patel_2017, Sayadi_2018, Sayadi_2019, Gao_2021, Sayadi_2021}) and Information Gain (Mutual Information) \cite{Peng_2005} (used by \cite{Singh_2017, Kwan_2020}). The Scikit-learn \cite {Scikit-learn} library for the Python and Weka \cite{Frank:2005aa} are tools frequently used in the \gls{HMD} field for \gls{FS}.

Since the number of events that can be potentially monitored exceeds the available \glspl{HPC}, some studies (for example, \cite{Patel_2017, Malone_2011, Khasawneh_2015}) also perform a preliminary manual \gls{FS} before data collection, thus reducing the number of software executions required to collect data. The selection is based on architectural and micro-architectural knowledge and other studies.

Eventually, in \gls{HMD}, \gls{ml} algorithms play a crucial role. Supervised and unsupervised learning techniques are employed in hardware-based malware detection. While for supervised detection, both benign and malignant samples, adequately annotated, are necessary, in unsupervised malware detection, the classifier is trained only with benign applications to perform anomaly detection \cite{Chandola_2009}. Unsupervised detection has two exciting advantages: (i) it does not require a malware dataset for training, and (ii) the classifier can detect zero-day malware~\cite{He:2021aa}. On the other side, unsupervised algorithms are complex, requiring more sophisticated analysis and resulting in complex hardware implementations.


Several traditional classification algorithm families are employed in \gls{HMD}: linear regression (LinearRegression and SimpleLinearRegression), logistic regression (Logistic and SimpleLogistic), Bayesian network (BayesNet and NaiveBayes), decision trees (J48 and REPTree), rule-based (JRIP, OneR and PART), \gls{ann} (MultiLayerPerceptron), \gls{KNN} (IBk), ensemble learning (AdaBoostM1, Bagging and RandomForest) and \gls{SVM} (SMO) \cite{Goodfellow_2016}. The algorithms in parentheses refer to specific Weka implementations, which are commonly used in the context of \gls{HMD}. Further details on these families and their implementations in Weka can be found in \cite{Frank:2005aa}.

Eventually, a crucial consideration is the trade-off between monitoring more events for better application characterization and detector performance and the impact on runtime applicability. Some studies used many events, exceeding available \glspl{HPC}, necessitating multiple application runs \cite{Demme_2013, Singh_2017, Sayadi_2017}. This trade-off is further addressed in \gls{ml} solutions discussed in Section \ref{subsec:machine_learning_techniques}.
\section{Hardware-based detection Assessment}
\label{sec:hw-based_performance_efficiency}

The following sections analyze the performance and efficiency of the state-of-the-art in \gls{HMD} and explore \gls{ml} techniques to enhance detector performance.


\subsection{Performance}
\label{subsubsec:performance}

Tables \ref{tab:performance_analysis} and \ref{tab:main_studies} provide a comprehensive overview of the literature contributions in the field, aiming to facilitate fair comparisons by presenting the best-case results in Table \ref{tab:performance_analysis}. Metrics were directly sourced from the paper's text whenever feasible, with manual extraction from reported \gls{ROC} curves employed only when necessary. The "Classification" column denotes the classification algorithm associated with the best result, with the Weka implementation serving as a reference. Conversely, Table \ref{tab:main_studies} outlines, for each contribution in Table \ref{tab:performance_analysis}, the range of considered scenarios in terms of malware, classifiers, and system characteristics. The values in Table \ref{tab:performance_analysis} underscore the efficacy of \gls{HMD} in supporting malware detection and highlight the overall high quality of the findings.
 
\begin{table*}
\begin{threeparttable}[b]
\caption{Summary of best-case performance from main studies in the hardware-based malware detection approach. \# \glspl{HPC} column refers to the number of hardware events the classifiers consider. Classification algorithm labels are based on Weka implementations used in the referenced studies. Evaluation metrics as defined in Section \ref{subsec:evaluation_metrics}: A is Accuracy, P is Precision, S is Specificity, and F1 is the F1-Score.}
\label{tab:performance_analysis}
\centering
\begin{tabularx}{\linewidth}
{
>{\hsize=.1\hsize\linewidth=\hsize}c
>{\hsize=.1\hsize\linewidth=\hsize}c
>{\hsize=.4\hsize\linewidth=\hsize}X
>{\hsize=.1\hsize\linewidth=\hsize}c
>{\hsize=.2\hsize\linewidth=\hsize}X
>{\hsize=.1\hsize\linewidth=\hsize}X
>{\hsize=.1\hsize\linewidth=\hsize}X
>{\hsize=.1\hsize\linewidth=\hsize}c
>{\hsize=.1\hsize\linewidth=\hsize}c
>{\hsize=.1\hsize\linewidth=\hsize}c
>{\hsize=.1\hsize\linewidth=\hsize}c
>{\hsize=.1\hsize\linewidth=\hsize}c
>{\hsize=.1\hsize\linewidth=\hsize}c
}
\toprule\
\textbf{Year} & \textbf{Ref.} & \textbf{Target} & \textbf{\# \glspl{HPC}} & \textbf{Classification} & \textbf{Learning} & \textbf{Latency} & \multicolumn{6}{c}{Evaluation Metrics} \\
 &  &  &  &  &  &  & \textbf{A} & \textbf{P} & \textbf{\gls{TPR}} & \textbf{S} & \textbf{F1} & \textbf{AUC}\\
\midrule\\
\rowcolor{shadecolor}
2013 & \cite{Demme_2013} & Android malware & 6 & Decision Tree & Offline
& NA & - & - & - & - & - & 0.83\\
\rowcolor{shadecolor}
& & Linux rootkits & 4 & KNN & Offline
& NA & - & - & 0.70\tnote{1} & - & - & -\\
\midrule
2014 & \cite{Tang_2014} & Internet Explorer exploitation & 4 & \gls{SVM} & Offline
& NA & - & - & - & - & - & 1.00\\
& & Adobe PDF Reader exploitation & 4 & \gls{SVM} & Offline
& NA & - & - & - & - & - & 1.00 \\
\midrule
\rowcolor{shadecolor}
2015 & \cite{Khasawneh_2015} & Ransomware & 5 & Logistic regression & Offline
& NA & - & - & - & - & - & 0.94 \\
\rowcolor{shadecolor}
& & Ransomware & 5  & Logistic regression (with Specialization) & Offline 
& NA & 0.87 & - & 0.81 & 0.96 & - & -\\
\midrule
2015 & \cite{Ozsoy_2015} & Viruses, worms, trojan horses, spyware, adware, and botnets & 5 & \gls{ann} & Offline
& NA & - & - & 1.00\tnote{1} & - & - & -\\
\midrule
\rowcolor{shadecolor}
2017 & \cite{Patel_2017} & Malware from various categories, sourced from VirusTotal~\cite{VirusTotal} dataset & 4 & BayesNet & Offline
& 0.624ms (SW) / 140ns (HW) & 0.85 & - & - & - & - & -\\
\midrule
2017 & \cite{Singh_2017} & Rootkits & 16 & \gls{SVM} &
& NA & 1.00 & 1.00 & 1.00 & - & 1.00 & - \\
\rowcolor{shadecolor}
\midrule
2018 & \cite{Sayadi_2018} & Malware from various categories, sourced from VirusTotal~\cite{VirusTotal} dataset & 4 & J48 (with Ensemble Learning) & Offline
& NA & 0.83 & - & - & - & - & 0.94 \\
\midrule
2018 & \cite{Zhou_2018} & Malware from various categories, sourced from VirusTotal~\cite{VirusTotal} dataset & 6 & Random Forest & Offline
& NA & - & 0.86 & 0.83 & - & 0.85 & 0.92\\
\midrule
\rowcolor{shadecolor}
2019 & \cite{Das_2019} & Malware from various categories, sourced from VX Heaven~\cite{Qiao:2016aa} dataset & 5 & J48 & Offline
& NA & - & 0.82 & 0.82 & - & 0.82 & 0.93 \\
\midrule
2019 & \cite{Sayadi_2019} & Backdoor & 4 & OneR & Offline
& NA & - & - & - & - & 0.94 & -\\
&  & Rookit & 4 & \gls{MLP} & Offline
& NA & - & - & - & - & 0.94 & -\\
&  & Virus & 4 & J48	 and ensemble learning (AdaBoostM1) & Offline
& NA & - & - & - & - & 0.96 & - \\
&  & Trojan & 4 & \gls{MLP} & Offline
& NA & - & - & - & - & 0.99 & -\\
\midrule
\rowcolor{shadecolor}
2021 & \cite{Gao_2021} & Trojan & 4 & JRIP  & Offline
& 20ns & - & 0.93 & - & - & - & -\\
\midrule
2021 & \cite{Sayadi_2021} & Stealthy rootkits & 4 & \gls{dnn} & Offline
& NA & 0.93 & 0.95 & 0.90 & - & 0.93 & 0.98 \\
\midrule
\rowcolor{shadecolor}
2022 & \cite{Torres_Liu_2022} & Data-only exploits~\cite{Hu_2015} & 50\tnote{2} & Two Classes-\gls{SVM} & Offline 
& NA & 0.99 & - & - & - & - & -\\
\rowcolor{shadecolor}
& & Data-only exploits~\cite{Hu_2015} & 6 & LZ78 & Offline
& NA & 0.84 & - & - & - & - & - \\
\midrule
2022 & \cite{Konstantinou:2022aa} & Stealthy attack on power grid & 6 & \gls{SVM} & Offline 
& 120s & 0.94 & - & - & - & - & - \\
\bottomrule\\
\end{tabularx}
\begin{tablenotes} [flushleft]
\item[1] Values extracted from ROC curves considering a false positive rate of 10\%.
\item[2] 50 is the whole set of features. This is why the authors also investigated a reduced set (in the following line).
\end{tablenotes}
\end{threeparttable}
\end{table*}

Among all contributions reported in Table~\ref{tab:performance_analysis}, authors in \cite{Konstantinou:2022aa} showcase the effectiveness of \gls{HMD} on real scenarios: DARPA \gls{RADICS}, Intel \gls{TDT}, and Microsoft Defender. This is a tangible exploitation of \gls{HMD} into actual products. Still, using a single type of classifier (i.e., \gls{SVM}) leaves room for research and improvements. 

\begin{table*}
\begin{threeparttable}[b]
\caption{Reference studies including details on the full list of targets and classifications approaches tested and details on the reference systems.}
\label{tab:main_studies}
\centering
\begin{tabularx}{\linewidth}{ccXXXX}
\toprule\
\textbf{Year} & \textbf{Ref.} & \textbf{Targets} & \textbf{Classification} & \textbf{Devices} & \textbf{OS}\\
\midrule\\
\rowcolor{shadecolor}
2013 & \cite{Demme_2013} & Android malware, Linux rootkits & Decision trees, \gls{ann}, \gls{KNN}, Random Forest & Arm Cortex-A9 OMAP4460, Intel Xeon X5550 & Android 4.1.1-1 (kernel 3.2), Linux kernel 2.6.32\\
\midrule
2014 & \cite{Tang_2014} & Exploitations on Internet Explorer and Adobe PDF Reader  & \gls{SVM} & Intel IvyBridge Core i7 & Windows XP\\
\midrule
\rowcolor{shadecolor}
2015 & \cite{Khasawneh_2015} & Ransomware, password stealers, trojan horses, backdoor, worms & Logistic regression	(w/o specialization) & Not specified & Windows 7\\
\midrule
2015 & \cite{Ozsoy_2015} & Viruses, worms, trojan horses, spyware, adware, and botnets & \gls{ann} &
Not specified, Altera EP4CE115 & Windows 7\\
\midrule
\rowcolor{shadecolor}
2017 & \cite{Patel_2017} & Malware from various categories, sourced from VirusTotal~\cite{VirusTotal} dataset & Logistic, SimpleLogistic, BayesNet, NaiveBayes, J48, PART, JRIP, OneR, MultiLayerPerceptron, SMO, SGD & Intel Haswell Core i5-4590, Xilinx Virtex 7 & Ubuntu 14.04 (kernel 4.4)\\
\midrule
2017 & \cite{Singh_2017} & Rootkits & \gls{SVM}, Decision tree, OC-SVM, Naive Bayes & Intel IvyBridge and Broadwell & Windows 7\\
\midrule
\rowcolor{shadecolor}
2018 & \cite{Sayadi_2018} & Malware from various categories, sourced from VirusTotal~\cite{VirusTotal} dataset & BayesNet, J48, REPTree, JRIP, OneR, MultiLayerPerceptron, SMO, SGD (w/o ensemble learning based on AdaBoostM1, Bagging) & Intel Xeon X5550, Xilinx Virtex 7 & Ubuntu 14.04 (kernel 4.4)\\
\midrule
2018 & \cite{Zhou_2018} & Malware from various categories, sourced from VirusTotal~\cite{VirusTotal} dataset & Decision trees, Naive Bayes, ANN, KNN, Random Forest (w/o ensemble learning AdaBoost) & AMD Bulldozer & Windows 7\\
\midrule
\rowcolor{shadecolor}
2019 & \cite{Das_2019} & Malware from various categories, sourced from VX Heaven~\cite{Qiao:2016aa} dataset & J48, IBk, SMO & Intel Sandy Bridge, Haswell, and Skylake & Ubuntu 16.04\\
\midrule
2019 & \cite{Sayadi_2019} & Backdoor, rootkits, viruses, trojan horses & J48, JRIP, OneR, \gls{MLP} (w/o AdaBoostM1) & Intel Xeon X5550, Xilinx Virtex 7 & Ubuntu 14.04 (kernel 4.4)\\
\midrule
\rowcolor{shadecolor}
2021 & \cite{Gao_2021} & Worms, rootkits, viruses, trojan horses & REPTree, JRIP, OneR, \gls{MLP}, SGD & Intel Xeon X5550, Xilinx Virtex 7 & Ubuntu 14.04 (kernel 4.4)\\
\midrule
2021 & \cite{Sayadi_2021} & Stealthy backdoor, rootkits, and trojan horses & \gls{dnn} &Intel Xeon X5550 &
Ubuntu 14.04 (kernel 4.4)\\
\midrule
\rowcolor{shadecolor}
2022 & \cite{Torres_Liu_2022} & Data-only exploitation & TC-SVM, OC-SVM, LZ78 & Intel Nehalem Core i7-920 & Ubuntu 16.04 (kernel 4.13)\\
\midrule
2022 & \cite{Konstantinou:2022aa} & Stealthy attack on power grid &\gls{SVM} & OpenPLC controller with Raspberry PI & 8-bus power grid in a PowerWorld simulator \\
\bottomrule\\
\end{tabularx}
\end{threeparttable}
\end{table*}

As most of the current works on \gls{HMD} rely on \gls{ml} classifiers, the analysis conducted by Patel et al. \cite{Patel_2017}, summarized in Table \ref{tab:classifiers_analysis}, is particularly interesting. The authors thoroughly analyze eleven \gls{ml} classification algorithms (based on Weka\cite{Frank:2005aa} implementations). The goal was to understand the trade-offs between the design parameters offered by the algorithms. The chosen metric to evaluate performance was accuracy. The dataset used for training and testing the algorithms was extracted using the PERF tool in intervals of 10 ms executed in an Intel Haswell Core i5-4590 processor running Ubuntu 14.04 with Linux kernel 4.4. The baseline of benign application comprises the Mibench benchmark suite \cite{Guthaus_2001}, Linux system programs, browsers, text editors, and word processors. The malware came from the VirusTotal dataset. Since the \glspl{HPC} available in an Intel architecture are considerable, the accuracy of \gls{ml} algorithms covers different numbers (i.e., 32, 8, 4, 2, and 1) of hardware events. Table \ref{tab:classifiers_analysis} reports the accuracy for 4 hardware events, a reasonable quantity for concurrent monitoring in most modern processors, even in embedded scenarios~\cite{Dutto:2021aa}. JRIP (rule-based) presented the top accuracy, followed by four classifiers with the same top-two accuracy: J48 (decision-tree), OneR and PART (rule-based), and SGD. In this case, most classifiers have accuracy above 80\%. Another interesting observation is that reducing the hardware events below four significantly impacts the performance of most classifiers.

Similar findings are reported in Torres and Liu \cite{Torres_Liu_2022}. While the authors concentrated on a particular malware subclass (data-only exploits from \cite{Hu_2015}), they implemented two different experiments on different classifiers, distinguishing between using the complete set of 50 features or a smaller set of 6 features. The findings report a very high accuracy on the complete set of features (as seen on the first of the two rows dedicated to the paper in Table \ref{tab:performance_analysis}) and a degradation when only a subset is used.

\begin{table}
\begin{threeparttable}[b]
\caption{Performance and efficiency of classifiers based on Weka implementations. Extracted from \cite{Patel_2017}.}
\label{tab:classifiers_analysis}
\centering
\begin{tabularx}{\linewidth}{
>{\hsize=.5\hsize\linewidth=\hsize}l
>{\hsize=.1\hsize\linewidth=\hsize}c
>{\hsize=.1\hsize\linewidth=\hsize}c
>{\hsize=.1\hsize\linewidth=\hsize}c
>{\hsize=.1\hsize\linewidth=\hsize}c
>{\hsize=.1\hsize\linewidth=\hsize}c
}
\toprule\
\multirow{3}{*}{Classifier} & \multirow{3}{*}{Accuracy} & SW  & \multicolumn{3}{c}{HW}                \\ 
                            &                           & Latency  & Latency  & Power  & Area\tnote{1} \\ 
                            &                           &    (ms)          &    (ns)                & (W) &       \\
 \midrule\\
\rowcolor{shadecolor}
BayesNet & 81.13 & 0.624 & 140 & 0.44 & 6794 \\
\midrule
J48 & 82.07 & 0.663 & 60 & 0.44 & 1801 \\
\midrule
\rowcolor{shadecolor}
JRIP & 83.96 & 0.653 & 90 & 0.44 & 1504 \\
\midrule
Logistic & 79.24 & 0.844 & 340 & 0.63 & 13041 \\
\midrule
\rowcolor{shadecolor}
\gls{MLP} & 81.13 & 0.870 & 40 & 1.03 & 36252 \\
\midrule
NaiveBayes & 78.30 & 0.802 & 10 & 1.34 & 58177 \\
\midrule
\rowcolor{shadecolor}
OneR & 82.07 & 0.653 & 220 & 0.32 & 1258 \\
\midrule
PART & 82.07 & 0.642 & 680 & 0.44 & 2131 \\
\midrule
\rowcolor{shadecolor}
SGD & 82.07 & 0.652 & 340 & 0.44 & 2556 \\
\midrule
SimpleLogistic & 79.24 & 0.648 & 3020 & 0.45 & 4721 \\
\midrule
\rowcolor{shadecolor}
SMO & 73.58 & 0.652 & 2330 & 0.44 & 2556 \\
\bottomrule\\
\end{tabularx}
\begin{tablenotes} [flushleft]
\item[1] The area is a function of total lookup tables, flip-flops, and DSP blocks.
\end{tablenotes}
\end{threeparttable}
\end{table}

\subsection{Efficiency}
\label{subsec:efficiency}

Alongside the detection quality, the \gls{HMD} aims to reduce the detectors cost in terms of resources. As the data required for the classification come from the hardware layer of the system stack, most studies evaluate FPGA-based implementations of \gls{ml} classifiers, providing measures for the power consumption and the area as the goal is to understand the trade-offs between the design parameters offered by the algorithms. When the classifier is software-based, the evaluation usually includes the latency, avoiding further monitoring of other resources. Unfortunately, as seen in Table~\ref{tab:performance_analysis}, not all works report the latency of the detection or, more in general, the costs of it. Generally, whenever the detection is performed at the software level, the latency is less than 1 ms. At the same time, more optimized hardware implementations can scale down to tens or hundreds of ns.

As reported in the previous section, the work from Patel et al. \cite{Patel_2017} covered a thorough analysis and, for this reason, is undoubtedly an excellent candidate to show the efficiency of the methodology. For hardware implementation, authors used the Xilinx Virtex 7 FPGA, implemented Weka models in C code, and used the Xilinx \gls{HLS} compiler to generate the final bitstream. The latency was evaluated both in software and hardware implementations. Authors implemented the classification algorithms in software at the \gls{OS} kernel level, which includes the time to read the \gls{HPC} and execute the classifiers. Eventually, the Intel Turbo Boost technology was disabled, as it might introduce errors in the time measurement, and the CPU governor was operating at a constant frequency of 800 MHz. The IP cores with the algorithms were synthesized in Vivado to estimate the power consumption, considering a 100 MHz clock. Power estimation contains both static power and dynamic power consumption of digital logic. 

Values in Table \ref{tab:classifiers_analysis} show the considerable difference between the latencies in software and hardware implementations. Software implementations have latencies almost in the order of milliseconds (ranging from 0.624ms to 0.870ms, best and worst cases). In contrast, hardware implementations are in the order of nanoseconds (ranging, in this case, from 10ns to 3020ns). The authors underlined that these slow profiles displayed by classifiers in the kernel space are three orders bigger than several malware executions (ranging in microseconds). Other findings related to latency are crucial to highlight. In software implementations, the latency for reading the \gls{HPC} is negligible when monitoring a single core but may increase significantly when monitoring multiple cores.
Moreover, the more \glspl{HPC} to read, the longer it takes. Concerning the classification algorithms, BayesNet (Bayesian network), PART (rule-based), and SimpleLogistic (logistic regression) showed the lowest latency values when implemented in software. Conversely, none of these three are on the list of the top three low latencies in hardware. NaiveBayes (Bayesian network), \gls{MLP} (\gls{ann}), and J48 (decision tree) are the three best hardware implementations. This paradox demonstrates the uncorrelation between the algorithms' latencies when comparing implementations at the kernel space and hardware.

\subsection{Machine learning techniques considerations}
\label{subsec:machine_learning_techniques}

Recent studies have explored various \gls{ml} methods to enhance the performance of \gls{HMD} detection approaches, especially in the last five years. These techniques aim to overcome the challenge of limited application characterization due to the concurrent capacity of \glspl{PMU} to monitor hardware events. While these methods show performance improvements, they often introduce increased complexity in classifiers, resulting in reduced efficiency, i.e., higher power consumption and increased area requirements. This section discusses ensemble learning, specialization, adaptive detection, and time series \gls{ml} techniques in \gls{HMD}.

In ensemble learning, multiple \gls{ml} algorithms are trained separately to create a classifier, combining their results to improve decision accuracy \cite{Dietterich_2000}. In \gls{HMD}, ensemble classifiers leverage the characteristics of individual algorithms to detect various types of malware while minimizing hardware events for runtime detection \cite{Khasawneh_2015, Sayadi_2018, Sayadi_2019}. However, the performance gains come with increased complexity and efficiency overhead \cite{Sayadi_2018, Gao_2021}.

Sayadi et al. \cite{Sayadi_2018} assessed the efficiency impact of ensemble learning in a malware detector on Xilinx Virtex 7 FPGA. Significant latency increases were observed when comparing a general classifier with 8 \glspl{HPC} to a Boosted classifier~\cite{Freund_Schapire_1997} with 4 \glspl{HPC}. When Boosted, the general \gls{MLP} algorithm passed from a latency of 3020ns to a latency of 5910ns. OneR increased from 10ns to 700ns, and J48  increased from 90ns to 670ns. In terms of hardware cost, the largest area increases were observed in OneR (from 2.1\% to 5.1\%), JRIP (from 2.5\% to 5.3\%), and BayesNet (from 11.5\% to 13.6\%). Conversely, J48, REPTree, and \gls{MLP} showed smaller area increases. The findings highlight substantial overhead in both latency and hardware costs. 

Another interesting \gls{ml} technique is the specialization. Instead of training a single multi-class classifier able to recognize several malware categories, different classifiers are trained, each specialized in detecting a specific malware. Authors in \cite{Khasawneh_2015} discuss and explore specialized detectors in \gls{HMD}. They used a logistic regression-based classifier for each malware class. As a result, the proposed detectors reduced the false positive rate by more than half compared to a single detector while increasing the detection rate. The authors proposed a two-level detector in the same paper, mixing a first level based on the hardware detection approach and a second level based on the software detection approach. The hardware detector was based on specialized ensemble techniques. The latency of this scheme was compared with malware detection purely based on software methods. As a result, they reported average latency reduced to 1/6.6 when the fraction of malware is low and latency reduced to 1/3.1 when 20\% of the programs are malware.

In 2019, Sayadi et al. \cite{Sayadi_2019} introduced a specialized two-stage malware detector, leveraging ensemble learning techniques, significantly improving accuracy. The first stage classifies applications into benign or malware categories (virus, rootkit, backdoor, and trojan horse). The second stage deploys an \gls{ml} classification algorithm that works best for each category of malware. Their 2021 work \cite{Sayadi_2021} continued using specialization for an accurate and run-time stealthy malware detector. They also evaluated the efficiency overhead of their specialized and ensemble learning malware detector, implemented on Xilinx Virtex 7 FPGA. A comparison of a general classifier with 4 \glspl{HPC} to a Boosted classifier with 4 \glspl{HPC} revealed notable latency increases for \gls{MLP} (from 1.020 to 5.910 ms), OneR (from 10 to 700 ns), J48 (from 30 to 670 ns), and JRIP (from 20 to 560 ns). \gls{MLP} (from 43.2\% to 61.7\%), JRIP (from 0.26\% to 5.3\%), OneR (from 0.49\% to 5.1\%) and J48 (from 0.93\% to 4.3\%) exhibited considerable increases regarding hardware cost. The findings emphasize substantial latency and hardware cost overhead.

Adaptive detection was proposed by Gao et al. \cite{Gao_2021} to optimize the performance versus cost. It targets higher or similar performance as ensemble learning, with a reduced cost. The technique leverages the concept that the \gls{ml} algorithm employed in the detector strongly correlates both the nature of the scrutinized malware and the overall performance metric. Adaptive detection involves a dynamic framework that assesses all underlying \gls{ml} algorithms in real time, opting for the optimal classifier to identify malicious patterns effectively. The implementation encompasses two primary online stages: (i) algorithm selection and (ii) malware detection. Consequently, only the most efficient ML-based detector is employed to differentiate malware from the benign class, eliminating the need to acquire results from individual base detectors and enhancing overall efficiency.

In the adaptive detector proposed by Gao et al. \cite{Gao_2021}, the algorithm selection step is done by a lightweight tree-based decision-making algorithm that accurately selects the most efficient model for inference. As a result, the scheme showed up to a 94\% detection rate while improving the cost-efficiency by more than 5X compared to existing ensemble-based malware detection methods.


Eventually, time series classification is fundamental to understanding the key concept behind hardware-based malware detection. The intuition driving this technique stems from the program's phase behavior, transforming malware detection into a time series classification problem. In addressing this challenge, Sayadi et al., as outlined in \cite{Sayadi_2020} and \cite{Sayadi_2021}, introduced a time series \gls{ml} technique designed to identify stealthy malware in real time. In scenarios where attackers embed malicious files within benign programs on target hosts, executing both applications as a single thread, traditional signature-based antivirus tools falter. Embedded malware remains elusive even when the exact malware signature is in the detector database. The authors proposed a classifier based on a \glspl{FCNN} and exclusively utilized branch instructions as a low-level feature in their solution. The results demonstrated the efficacy of their technique, achieving a remarkable average detection performance of 94\% with only one \gls{HPC} feature, surpassing state-of-the-art detection methods. This enhanced performance, however, comes at a higher computational cost associated with employing a deep-learning-based solution.

While not explicitly implementing a time series technique, also \cite{Konstantinou:2022aa} reports similar results on the Intel \gls{TDT} use case. Although no specific numbers are provided, the paper compares the \gls{FFT} counting traces of the branch instructions and branch misprediction events for the WannaCry ransomware, underlining the significant difference with or without the ransomware. 

\section{Conclusions and research challenges}
\label{sec:conclusion_challenges}

In summary, this paper provided a comprehensive overview of \gls{HMD} field, with a detailed analysis of hardware-based detection, harnessing the power of \glspl{HPC} and \gls{ml}. The advantages of \gls{HMD} include resilience to malware subverting the protection mechanism, adaptability to code variants and unknown malware, low complexity and overhead, potential for run-time detection, and cost reduction.

However, challenges persist in \gls{HMD}. The detection accuracy is the most significant challenge as classifiers have a statistical nature. Thus, their results are not deterministic, and ongoing research aims to minimize errors by exploring complex classifiers. In cases where high accuracy is unattainable, a potential solution combines software and hardware-based detectors concurrently, with hardware as the primary defense.
Moreover, ensuring consistency, accuracy, and standardization of hardware monitoring units (including \glspl{HPC}) is crucial for trustworthiness. Chip manufacturers can contribute by designing appropriate modules and providing comprehensive documentation. The limited number of \glspl{HPC} in mobile and \gls{iot} devices poses a feasibility challenge for this approach in these domains. Addressing these challenges will contribute to the continued advancement and effectiveness of \gls{HMD}.

\bibliographystyle{IEEEtran}
\bibliography{references}

\begin{thebibliography}{10}
\providecommand{\url}[1]{#1}
\csname url@samestyle\endcsname
\providecommand{\newblock}{\relax}
\providecommand{\bibinfo}[2]{#2}
\providecommand{\BIBentrySTDinterwordspacing}{\spaceskip=0pt\relax}
\providecommand{\BIBentryALTinterwordstretchfactor}{4}
\providecommand{\BIBentryALTinterwordspacing}{\spaceskip=\fontdimen2\font plus
\BIBentryALTinterwordstretchfactor\fontdimen3\font minus \fontdimen4\font\relax}
\providecommand{\BIBforeignlanguage}[2]{{%
\expandafter\ifx\csname l@#1\endcsname\relax
\typeout{** WARNING: IEEEtran.bst: No hyphenation pattern has been}%
\typeout{** loaded for the language `#1'. Using the pattern for}%
\typeout{** the default language instead.}%
\else
\language=\csname l@#1\endcsname
\fi
#2}}
\providecommand{\BIBdecl}{\relax}
\BIBdecl

\bibitem{McGraw_Morrisett_2000}
G.~McGraw and G.~Morrisett, ``Attacking malicious code: A report to the infosec research council,'' \emph{IEEE Software}, vol.~17, no.~5, pp. 33--41, 2000.

\bibitem{Alsmadi:2021aa}
T.~Alsmadi and N.~Alqudah, ``A survey on malware detection techniques,'' in \emph{2021 International Conference on Information Technology (ICIT)}, 2021, pp. 371--376.

\bibitem{Stallings_Brown_2014}
W.~Stallings and L.~Brown, \emph{Computer Security: Principles and Practice}, 3rd~ed.\hskip 1em plus 0.5em minus 0.4em\relax USA: Prentice Hall Press, 2014.

\bibitem{SonicWall_2020}
SonicWall, ``New sonicwall research finds aggressive growth in ransomware, rise in iot attacks,'' Available at \url{https://www.sonicwall.com/news/new-sonicwall-research-finds-aggressive-growth-in-ransomware-rise-in-iot-attacks/} (Accessed in September 18, 2023).

\bibitem{GlobalRisckReport2023}
\BIBentryALTinterwordspacing
``Global risk report 2023,'' Jan 2023. [Online]. Available: \url{https://www.weforum.org/publications/global-risks-report-2023/in-full/}
\BIBentrySTDinterwordspacing

\bibitem{cybersecurity_almanac_2023}
Cisco and C.~Ventures, ``2023 cybersecurity almanac: 100 facts, figures, predictions and statistics,'' Available at \url{http://cybersecurityventures.com/cybersecurity-almanac-2023/} (Accessed in September 18, 2023).

\bibitem{Ye:2017aa}
\BIBentryALTinterwordspacing
Y.~Ye, T.~Li, D.~Adjeroh, and S.~S. Iyengar, ``A survey on malware detection using data mining techniques,'' \emph{ACM Computing Surveys}, vol.~50, no.~3, pp. 1--40, Jun. 2017. [Online]. Available: \url{http://dx.doi.org/10.1145/3073559}
\BIBentrySTDinterwordspacing

\bibitem{Ucci:2019aa}
\BIBentryALTinterwordspacing
D.~Ucci, L.~Aniello, and R.~Baldoni, ``Survey of machine learning techniques for malware analysis,'' \emph{Computers \& Security}, vol.~81, pp. 123--147, 2019. [Online]. Available: \url{https://www.sciencedirect.com/science/article/pii/S0167404818303808}
\BIBentrySTDinterwordspacing

\bibitem{Gibert:2020aa}
\BIBentryALTinterwordspacing
D.~Gibert, C.~Mateu, and J.~Planes, ``The rise of machine learning for detection and classification of malware: Research developments, trends and challenges,'' \emph{Journal of Network and Computer Applications}, vol. 153, p. 102526, 2020. [Online]. Available: \url{https://www.sciencedirect.com/science/article/pii/S1084804519303868}
\BIBentrySTDinterwordspacing

\bibitem{Qiu:2020aa}
\BIBentryALTinterwordspacing
J.~Qiu, J.~Zhang, W.~Luo, L.~Pan, S.~Nepal, and Y.~Xiang, ``A survey of android malware detection with deep neural models,'' \emph{ACM Comput. Surv.}, vol.~53, no.~6, dec 2020. [Online]. Available: \url{https://doi.org/10.1145/3417978}
\BIBentrySTDinterwordspacing

\bibitem{Liu:2022aa}
\BIBentryALTinterwordspacing
Y.~Liu, C.~Tantithamthavorn, L.~Li, and Y.~Liu, ``Deep learning for android malware defenses: A systematic literature review,'' \emph{ACM Comput. Surv.}, vol.~55, no.~8, dec 2022. [Online]. Available: \url{https://doi.org/10.1145/3544968}
\BIBentrySTDinterwordspacing

\bibitem{Catal:2022aa}
\BIBentryALTinterwordspacing
C.~Catal, G.~Giray, and B.~Tekinerdogan, ``Applications of deep learning for mobile malware detection: A systematic literature review,'' \emph{Neural Computing and Applications}, vol.~34, no.~2, pp. 1007--1032, Jan 2022. [Online]. Available: \url{https://doi.org/10.1007/s00521-021-06597-0}
\BIBentrySTDinterwordspacing

\bibitem{Deldar:2023aa}
\BIBentryALTinterwordspacing
F.~Deldar and M.~Abadi, ``Deep learning for zero-day malware detection and classification: A survey,'' \emph{ACM Comput. Surv.}, vol.~56, no.~2, sep 2023. [Online]. Available: \url{https://doi.org/10.1145/3605775}
\BIBentrySTDinterwordspacing

\bibitem{Malone_2011}
\BIBentryALTinterwordspacing
C.~Malone, M.~Zahran, and R.~Karri, ``Are hardware performance counters a cost effective way for integrity checking of programs,'' in \emph{Proceedings of the Sixth ACM Workshop on Scalable Trusted Computing}, ser. STC '11.\hskip 1em plus 0.5em minus 0.4em\relax New York, NY, USA: Association for Computing Machinery, 2011, pp. 71--76. [Online]. Available: \url{https://doi.org/10.1145/2046582.2046596}
\BIBentrySTDinterwordspacing

\bibitem{Demme_2013}
\BIBentryALTinterwordspacing
J.~Demme, M.~Maycock, J.~Schmitz, A.~Tang, A.~Waksman, S.~Sethumadhavan, and S.~Stolfo, ``On the feasibility of online malware detection with performance counters,'' \emph{SIGARCH Comput. Archit. News}, vol.~41, no.~3, pp. 559--570, jun 2013. [Online]. Available: \url{https://doi.org/10.1145/2508148.2485970}
\BIBentrySTDinterwordspacing

\bibitem{Sherwood_2003}
T.~Sherwood, E.~Perelman, G.~Hamerly, S.~Sair, and B.~Calder, ``Discovering and exploiting program phases,'' \emph{IEEE Micro}, vol.~23, no.~6, pp. 84--93, 2003.

\bibitem{Isci_2006}
C.~Isci, G.~Contreras, and M.~Martonosi, ``Live, runtime phase monitoring and prediction on real systems with application to dynamic power management,'' in \emph{2006 39th Annual IEEE/ACM International Symposium on Microarchitecture (MICRO'06)}, 2006, pp. 359--370.

\bibitem{He:2021aa}
Z.~He, T.~Miari, H.~M. Makrani, M.~Aliasgari, H.~Homayoun, and H.~Sayadi, ``When machine learning meets hardware cybersecurity: Delving into accurate zero-day malware detection,'' in \emph{2021 22nd International Symposium on Quality Electronic Design (ISQED)}, 2021, pp. 85--90.

\bibitem{Christodorescu_2007}
M.~Christodorescu, S.~Jha, D.~Maughan, D.~Song, and C.~Wang, \emph{Advances in Information Security: Malware detection}, 1st~ed.\hskip 1em plus 0.5em minus 0.4em\relax New York, NY: Springer, 2007.

\bibitem{NIST_Glossary_2023}
NIST, ``Glossary,'' Available at \url{https://csrc.nist.gov/glossary} (Accessed in March 07, 2023).

\bibitem{ENISA_Botnets_2023}
ENISA, ``Botnets,'' Available at \url{https://www.enisa.europa.eu/topics/incident-response/glossary/botnets} (Accessed in March 07, 2023).

\bibitem{Aycock_2010}
J.~Aycock, \emph{Computer Viruses and Malware}, ser. Advances in Information Security.\hskip 1em plus 0.5em minus 0.4em\relax Springer US, 2006.

\bibitem{You_Yim_2010}
I.~You and K.~Yim, ``Malware obfuscation techniques: A brief survey,'' in \emph{2010 International Conference on Broadband, Wireless Computing, Communication and Applications}, 2010, pp. 297--300.

\bibitem{Rad_2012}
B.~Bashari~Rad, M.~Masrom, and S.~Ibrahim, ``Camouflage in malware: From encryption to metamorphism,'' \emph{International Journal of Computer Science And Network Security (IJCSNS)}, vol.~12, pp. 74--83, 01 2012.

\bibitem{Stolfo_2007}
S.~J. Stolfo, K.~Wang, and W.-J. Li, ``Towards stealthy malware detection,'' in \emph{Malware Detection}, M.~Christodorescu, S.~Jha, D.~Maughan, D.~Song, and C.~Wang, Eds.\hskip 1em plus 0.5em minus 0.4em\relax Boston, MA: Springer US, 2007, pp. 231--249.

\bibitem{Rudd_2017}
E.~M. Rudd, A.~Rozsa, M.~G{\"u}nther, and T.~E. Boult, ``A survey of stealth malware attacks, mitigation measures, and steps toward autonomous open world solutions,'' \emph{IEEE Communications Surveys \& Tutorials}, vol.~19, no.~2, pp. 1145--1172, 2017.

\bibitem{Nadim:2021aa}
M.~Nadim, D.~Akopian, and W.~Lee, ``A review on learning-based detection approaches of the kernel-level rootkit,'' in \emph{2021 International Conference on Engineering and Emerging Technologies (ICEET)}, 2021, pp. 1--6.

\bibitem{Wong_Stamp_2006}
W.~Wong and M.~Stamp, ``Hunting for metamorphic engines,'' \emph{Journal in Computer Virology}, vol.~2, pp. 211--229, 11 2006.

\bibitem{Konstantinou_2008}
E.~Konstantinou, ``Metamorphic virus: Analysis and detection,'' Ph.D. dissertation, University of London, London UK, 2008.

\bibitem{Brezinski_2023}
\BIBentryALTinterwordspacing
K.~Brezinski, K.~Ferens, and K.~Rantos, ``Metamorphic malware and obfuscation: A survey of techniques, variants, and generation kits,'' \emph{Sec. and Commun. Netw.}, vol. 2023, sep 2023. [Online]. Available: \url{https://doi.org/10.1155/2023/8227751}
\BIBentrySTDinterwordspacing

\bibitem{Van_der_Veen_2012}
V.~van~der Veen, N.~dutt Sharma, L.~Cavallaro, and H.~Bos, ``Memory errors: The past, the present, and the future,'' in \emph{Research in Attacks, Intrusions, and Defenses}, D.~Balzarotti, S.~J. Stolfo, and M.~Cova, Eds.\hskip 1em plus 0.5em minus 0.4em\relax Berlin, Heidelberg: Springer Berlin Heidelberg, 2012, pp. 86--106.

\bibitem{Khasawneh_2015}
\BIBentryALTinterwordspacing
K.~N. Khasawneh, M.~Ozsoy, C.~Donovick, N.~Abu-Ghazaleh, and D.~Ponomarev, ``Ensemble learning for low-level hardware-supported malware detection,'' in \emph{Proceedings of the 18th International Symposium on Research in Attacks, Intrusions, and Defenses - Volume 9404}, ser. RAID 2015.\hskip 1em plus 0.5em minus 0.4em\relax Berlin, Heidelberg: Springer-Verlag, 2015, pp. 3--25. [Online]. Available: \url{https://doi.org/10.1007/978-3-319-26362-5\_1}
\BIBentrySTDinterwordspacing

\bibitem{Ozsoy_2015}
M.~Ozsoy, C.~Donovick, I.~Gorelik, N.~Abu-Ghazaleh, and D.~Ponomarev, ``Malware-aware processors: A framework for efficient online malware detection,'' in \emph{2015 IEEE 21st International Symposium on High Performance Computer Architecture (HPCA)}, 2015, pp. 651--661.

\bibitem{Tang_2014}
A.~Tang, S.~Sethumadhavan, and S.~J. Stolfo, ``Unsupervised anomaly-based malware detection using hardware features,'' in \emph{Research in Attacks, Intrusions and Defenses}, A.~Stavrou, H.~Bos, and G.~Portokalidis, Eds.\hskip 1em plus 0.5em minus 0.4em\relax Cham: Springer International Publishing, 2014, pp. 109--129.

\bibitem{Wang_Karri_2013}
X.~Wang and R.~Karri, ``Numchecker: Detecting kernel control-flow modifying rootkits by using hardware performance counters,'' in \emph{2013 50th ACM/EDAC/IEEE Design Automation Conference (DAC)}, 2013, pp. 1--7.

\bibitem{Ray_Ligatti_2012}
\BIBentryALTinterwordspacing
D.~Ray and J.~Ligatti, ``Defining code-injection attacks,'' \emph{SIGPLAN Not.}, vol.~47, no.~1, pp. 179--190, jan 2012. [Online]. Available: \url{https://doi.org/10.1145/2103621.2103678}
\BIBentrySTDinterwordspacing

\bibitem{Prandini_Ramilli_2012}
M.~Prandini and M.~Ramilli, ``Return-oriented programming,'' \emph{IEEE Security \& Privacy}, vol.~10, no.~6, pp. 84--87, 2012.

\bibitem{Bletsch_2011}
\BIBentryALTinterwordspacing
T.~Bletsch, X.~Jiang, V.~W. Freeh, and Z.~Liang, ``Jump-oriented programming: a new class of code-reuse attack,'' in \emph{Proceedings of the 6th ACM Symposium on Information, Computer and Communications Security}, ser. ASIACCS '11.\hskip 1em plus 0.5em minus 0.4em\relax New York, NY, USA: Association for Computing Machinery, 2011, pp. 30--40. [Online]. Available: \url{https://doi.org/10.1145/1966913.1966919}
\BIBentrySTDinterwordspacing

\bibitem{Chen_2005}
S.~Chen, J.~Xu, E.~C. Sezer, P.~Gauriar, and R.~K. Iyer, ``Non-control-data attacks are realistic threats,'' in \emph{Proceedings of the 14th Conference on USENIX Security Symposium - Volume 14}, ser. SSYM'05.\hskip 1em plus 0.5em minus 0.4em\relax USA: USENIX Association, 2005, p.~12.

\bibitem{Aslan_Samet_2020}
O.~A. Aslan and R.~Samet, ``A comprehensive review on malware detection approaches,'' \emph{IEEE Access}, vol.~8, pp. 6249--6271, 2020.

\bibitem{ISO_IEC_IEEE_Vocabulary_2017}
``Iso/iec/ieee international standard - systems and software engineering--vocabulary,'' pp. 1--541, 2017.

\bibitem{Sze_2020}
V.~Sze, Y.-H. Chen, T.-J. Yang, and J.~S. Emer, \emph{Efficient Processing of Deep Neural Networks}, ser. Synthesis Lectures on Computer Architecture Series.\hskip 1em plus 0.5em minus 0.4em\relax Cham, Switzerland: Springer Cham, 2020.

\bibitem{Idika_Mathur_2007}
\BIBentryALTinterwordspacing
N.~Idika and P.~Mathur, ``A survey of malware detection techniques,'' Purdue University, West Lafayette, USA, Tech. Rep., 2007. [Online]. Available: \url{https://api.semanticscholar.org/CorpusID:2216347}
\BIBentrySTDinterwordspacing

\bibitem{Alzarooni_2012}
K.~Alzarooni, ``Malware variant detection,'' Ph.D. dissertation, University College London, London, U.K., 2012.

\bibitem{Alonso:2023aa}
M.~Alonso, D.~Andreu, R.~Canal, S.~Di~Carlo, C.~Chenet, J.~Costa, A.~Girones, D.~Gizopoulos, V.~Karakostas, B.~Otero, G.~Papadimitriou, E.~Rodr{\'\i}guez, and A.~Savino, ``Validation, verification, and testing (vvt) of future risc-v powered cloud infrastructures: the vitamin-v horizon europe project perspective,'' in \emph{2023 IEEE European Test Symposium (ETS)}, 2023, pp. 1--6.

\bibitem{Sayadi_2018}
H.~Sayadi, N.~Patel, S.~M. P.D., A.~Sasan, S.~Rafatirad, and H.~Homayoun, ``Ensemble learning for effective run-time hardware-based malware detection: A comprehensive analysis and classification,'' in \emph{2018 55th ACM/ESDA/IEEE Design Automation Conference (DAC)}, 2018, pp. 1--6.

\bibitem{Sayadi_2019}
H.~Sayadi, H.~M. Makrani, S.~M. Pudukotai~Dinakarrao, T.~Mohsenin, A.~Sasan, S.~Rafatirad, and H.~Homayoun, ``2smart: A two-stage machine learning-based approach for run-time specialized hardware-assisted malware detection,'' in \emph{2019 Design, Automation \& Test in Europe Conference \& Exhibition (DATE)}, 2019, pp. 728--733.

\bibitem{Patel_2017}
N.~Patel, A.~Sasan, and H.~Homayoun, ``Analyzing hardware based malware detectors,'' in \emph{2017 54th ACM/EDAC/IEEE Design Automation Conference (DAC)}, 2017, pp. 1--6.

\bibitem{Sayadi_2022}
H.~Sayadi, M.~Aliasgari, F.~Aydin, S.~Potluri, A.~Aysu, J.~Edmonds, and S.~Tehranipoor, ``Towards ai-enabled hardware security: Challenges and opportunities,'' in \emph{2022 IEEE 28th International Symposium on On-Line Testing and Robust System Design (IOLTS)}, 2022, pp. 1--10.

\bibitem{Dutto:2021aa}
S.~Dutto, A.~Savino, and S.~Di~Carlo, ``Exploring deep learning for in-field fault detection in microprocessors,'' in \emph{2021 Design, Automation \& Test in Europe Conference \& Exhibition (DATE)}, 2021, pp. 1456--1459.

\bibitem{Torres_Liu_2022}
\BIBentryALTinterwordspacing
G.~Torres and C.~Liu, ``Where's waldo? identifying anomalous behavior of data-only attacks using hardware features,'' in \emph{Proceedings of the 19th ACM International Conference on Computing Frontiers}, ser. CF '22.\hskip 1em plus 0.5em minus 0.4em\relax New York, NY, USA: Association for Computing Machinery, 2022, pp. 75--84. [Online]. Available: \url{https://doi.org/10.1145/3528416.3530226}
\BIBentrySTDinterwordspacing

\bibitem{Kasap:2023aa}
D.~Kasap, A.~Carpegna, A.~Savino, and S.~Di~Carlo, ``Micro-architectural features as soft-error markers in embedded safety-critical systems: preliminary study,'' in \emph{2023 IEEE European Test Symposium (ETS)}, 2023, pp. 1--5.

\bibitem{Zhou_2018}
\BIBentryALTinterwordspacing
B.~Zhou, A.~Gupta, R.~Jahanshahi, M.~Egele, and A.~Joshi, ``Hardware performance counters can detect malware: Myth or fact?'' in \emph{Proceedings of the 2018 on Asia Conference on Computer and Communications Security}, ser. ASIACCS '18.\hskip 1em plus 0.5em minus 0.4em\relax New York, NY, USA: Association for Computing Machinery, 2018, pp. 457--468. [Online]. Available: \url{https://doi.org/10.1145/3196494.3196515}
\BIBentrySTDinterwordspacing

\bibitem{Zhou_2021}
------, ``A cautionary tale about detecting malware using hardware performance counters and machine learning,'' \emph{IEEE Design \& Test}, vol.~38, no.~3, pp. 39--50, 2021.

\bibitem{Botacin_Gregio_2022}
M.~Botacin and A.~Gr{\'e}gio, ``Why we need a theory of maliciousness: Hardware performance counters in security,'' in \emph{Information Security}, W.~Susilo, X.~Chen, F.~Guo, Y.~Zhang, and R.~Intan, Eds.\hskip 1em plus 0.5em minus 0.4em\relax Cham: Springer International Publishing, 2022, pp. 381--389.

\bibitem{Kim_2014}
\BIBentryALTinterwordspacing
Y.~Kim, R.~Daly, J.~Kim, C.~Fallin, J.~H. Lee, D.~Lee, C.~Wilkerson, K.~Lai, and O.~Mutlu, ``Flipping bits in memory without accessing them: an experimental study of dram disturbance errors,'' \emph{SIGARCH Comput. Archit. News}, vol.~42, no.~3, pp. 361--372, jun 2014. [Online]. Available: \url{https://doi.org/10.1145/2678373.2665726}
\BIBentrySTDinterwordspacing

\bibitem{Mutlu_2020}
\BIBentryALTinterwordspacing
O.~Mutlu and J.~S. Kim, ``Rowhammer: A retrospective,'' \emph{Trans. Comp.-Aided Des. Integ. Cir. Sys.}, vol.~39, no.~8, pp. 1555--1571, aug 2020. [Online]. Available: \url{https://doi.org/10.1109/TCAD.2019.2915318}
\BIBentrySTDinterwordspacing

\bibitem{Li_Gaudiot_2019}
C.~Li and J.-L. Gaudiot, ``Detecting malicious attacks exploiting hardware vulnerabilities using performance counters,'' in \emph{2019 IEEE 43rd Annual Computer Software and Applications Conference (COMPSAC)}, vol.~1, 2019, pp. 588--597.

\bibitem{Wang_Backer_2016}
X.~Wang and J.~Backer, ``Sigdrop: Signature-based rop detection using hardware performance counters,'' 2016.

\bibitem{NIST_CVE-2016-5195}
NIST, ``National vulnerability database: Cve-2016-5195 detail,'' Available at \url{http:// https://nvd.nist.gov/vuln/detail/cve-2016-5195} (2024/31/01).

\bibitem{Weaver_McKee_2008}
V.~M. Weaver and S.~A. McKee, ``Can hardware performance counters be trusted?'' in \emph{2008 IEEE International Symposium on Workload Characterization}, 2008, pp. 141--150.

\bibitem{Carelli:2018aa}
A.~Carelli, A.~Vallero, and S.~D. Carlo, ``Shielding performance monitor counters: a double edged weapon for safety and security,'' in \emph{2018 IEEE 24th International Symposium on On-Line Testing And Robust System Design (IOLTS)}, 2018, pp. 269--274.

\bibitem{Carelli:2019aa}
A.~Carelli, A.~Vallero, and S.~Di~Carlo, ``Performance monitor counters: Interplay between safety and security in complex cyber-physical systems,'' \emph{IEEE Transactions on Device and Materials Reliability}, vol.~19, no.~1, pp. 73--83, 2019.

\bibitem{Weaver_Terpstra_Moore_2013}
V.~M. Weaver, D.~Terpstra, and S.~Moore, ``Non-determinism and overcount on modern hardware performance counter implementations,'' in \emph{2013 IEEE International Symposium on Performance Analysis of Systems and Software (ISPASS)}, 2013, pp. 215--224.

\bibitem{Das_2019}
S.~Das, J.~Werner, M.~Antonakakis, M.~Polychronakis, and F.~Monrose, ``Sok: The challenges, pitfalls, and perils of using hardware performance counters for security,'' in \emph{2019 IEEE Symposium on Security and Privacy (SP)}, 2019, pp. 20--38.

\bibitem{Barrera_2020}
\BIBentryALTinterwordspacing
J.~Barrera, L.~Kosmidis, H.~Tabani, E.~Mezzetti, J.~Abella, M.~Fernandez, G.~Bernat, and F.~J. Cazorla, ``On the reliability of hardware event monitors in mpsocs for critical domains,'' in \emph{Proceedings of the 35th Annual ACM Symposium on Applied Computing}, ser. SAC '20.\hskip 1em plus 0.5em minus 0.4em\relax New York, NY, USA: Association for Computing Machinery, 2020, pp. 580--589. [Online]. Available: \url{https://doi.org/10.1145/3341105.3373955}
\BIBentrySTDinterwordspacing

\bibitem{Kadiyala_2020}
\BIBentryALTinterwordspacing
S.~P. Kadiyala, P.~Jadhav, S.-K. Lam, and T.~Srikanthan, ``Hardware performance counter-based fine-grained malware detection,'' \emph{ACM Trans. Embed. Comput. Syst.}, vol.~19, no.~5, sep 2020. [Online]. Available: \url{https://doi.org/10.1145/3403943}
\BIBentrySTDinterwordspacing

\bibitem{Ritter_2022}
M.~Ritter, A.~Tarraf, A.~Gei{\ss}, N.~Daoud, B.~Mohr, and F.~Wolf, ``Conquering noise with hardware counters on hpc systems,'' in \emph{2022 IEEE/ACM Workshop on Programming and Performance Visualization Tools (ProTools)}, 2022, pp. 1--10.

\bibitem{Sasongko_2023}
M.~A. Sasongko, M.~Chabbi, P.~H.~J. Kelly, and D.~Unat, ``Precise event sampling on amd versus intel: Quantitative and qualitative comparison,'' \emph{IEEE Transactions on Parallel and Distributed Systems}, vol.~34, no.~5, pp. 1594--1608, 2023.

\bibitem{Sprunt_2002}
B.~Sprunt, ``The basics of performance-monitoring hardware,'' \emph{IEEE Micro}, vol.~22, no.~4, pp. 64--71, 2002.

\bibitem{Doyle_2017}
N.~C. Doyle, E.~Matthews, G.~Holland, A.~Fedorova, and L.~Shannon, ``Performance impacts and limitations of hardware memory access trace collection,'' in \emph{Design, Automation \& Test in Europe Conference \& Exhibition (DATE), 2017}, 2017, pp. 506--511.

\bibitem{Abdulwahab_2022}
\BIBentryALTinterwordspacing
H.~M. Abdulwahab, S.~Ajitha, and M.~A.~N. Saif, ``Feature selection techniques in the context of big data: Taxonomy and analysis,'' \emph{Applied Intelligence}, vol.~52, no.~12, pp. 13\,568--13\,613, sep 2022. [Online]. Available: \url{https://doi.org/10.1007/s10489-021-03118-3}
\BIBentrySTDinterwordspacing

\bibitem{Chandrashekar_Sahin_2014}
\BIBentryALTinterwordspacing
G.~Chandrashekar and F.~Sahin, ``A survey on feature selection methods,'' \emph{Comput. Electr. Eng.}, vol.~40, no.~1, pp. 16--28, jan 2014. [Online]. Available: \url{https://doi.org/10.1016/j.compeleceng.2013.11.024}
\BIBentrySTDinterwordspacing

\bibitem{PAPI_1999}
S.~Browne, C.~Deane, G.~Ho, and P.~Mucci, ``Papi: A portable interface to hardware performance counters,'' in \emph{Proceedings of Department of Defense HPCMP Users Group Conference}, 1999-06 1999.

\bibitem{PERF_2009}
I.~Molnar and T.~Gleixner, ``Performance counters for linux,'' Available at \url{https:// https://lwn.net/Articles/337493} (2023/07/05).

\bibitem{Binkert_2011}
\BIBentryALTinterwordspacing
N.~Binkert, B.~Beckmann, G.~Black, S.~K. Reinhardt, A.~Saidi, A.~Basu, J.~Hestness, D.~R. Hower, T.~Krishna, S.~Sardashti, R.~Sen, K.~Sewell, M.~Shoaib, N.~Vaish, M.~D. Hill, and D.~A. Wood, ``The gem5 simulator,'' \emph{SIGARCH Comput. Archit. News}, vol.~39, no.~2, pp. 1--7, aug 2011. [Online]. Available: \url{https://doi.org/10.1145/2024716.2024718}
\BIBentrySTDinterwordspacing

\bibitem{Bruschi_2021}
N.~Bruschi, G.~Haugou, G.~Tagliavini, F.~Conti, L.~Benini, and D.~Rossi, ``Gvsoc: A highly configurable, fast and accurate full-platform simulator for risc-v based iot processors,'' in \emph{2021 IEEE 39th International Conference on Computer Design (ICCD)}, 2021, pp. 409--416.

\bibitem{Goodfellow_2016}
I.~Goodfellow, Y.~Bengio, and A.~Courville, \emph{Deep Learning}.\hskip 1em plus 0.5em minus 0.4em\relax MIT Press, 2016, \url{http://www.deeplearningbook.org}.

\bibitem{Gao_2021}
Y.~Gao, H.~M. Makrani, M.~Aliasgari, A.~Rezaei, J.~Lin, H.~Homayoun, and H.~Sayadi, ``Adaptive-hmd: Accurate and cost-efficient machine learning-driven malware detection using microarchitectural events,'' in \emph{2021 IEEE 27th International Symposium on On-Line Testing and Robust System Design (IOLTS)}, 2021, pp. 1--7.

\bibitem{Sayadi_2021}
\BIBentryALTinterwordspacing
H.~Sayadi, Y.~Gao, H.~Mohammadi~Makrani, J.~Lin, P.~C. Costa, S.~Rafatirad, and H.~Homayoun, ``Towards accurate run-time hardware-assisted stealthy malware detection: A lightweight, yet effective time series cnn-based approach,'' \emph{Cryptography}, vol.~5, no.~4, 2021. [Online]. Available: \url{https://www.mdpi.com/2410-387X/5/4/28}
\BIBentrySTDinterwordspacing

\bibitem{Duda_2000}
R.~O. Duda, P.~E. Hart, and D.~G. Stork, \emph{Pattern Classification (2nd Edition)}.\hskip 1em plus 0.5em minus 0.4em\relax USA: Wiley-Interscience, 2000.

\bibitem{Pearson_1895}
K.~Pearson, ``Note on regression and inheritance in the case of two parents,'' \emph{Proc. R. Soc. Lond.}, vol.~58, pp. 240--242, 1895.

\bibitem{Peng_2005}
H.~Peng, F.~Long, and C.~Ding, ``Feature selection based on mutual information criteria of max-dependency, max-relevance, and min-redundancy,'' \emph{IEEE Transactions on Pattern Analysis and Machine Intelligence}, vol.~27, no.~8, pp. 1226--1238, 2005.

\bibitem{Singh_2017}
\BIBentryALTinterwordspacing
B.~Singh, D.~Evtyushkin, J.~Elwell, R.~Riley, and I.~Cervesato, ``On the detection of kernel-level rootkits using hardware performance counters,'' in \emph{Proceedings of the 2017 ACM on Asia Conference on Computer and Communications Security}, ser. ASIA CCS '17.\hskip 1em plus 0.5em minus 0.4em\relax New York, NY, USA: Association for Computing Machinery, 2017, pp. 483--493. [Online]. Available: \url{https://doi.org/10.1145/3052973.3052999}
\BIBentrySTDinterwordspacing

\bibitem{Kwan_2020}
A.~Kwan, ``Malware detection at the microarchitecture level using machine learning techniques,'' Available at \url{http:// https://scholarworks.calstate.edu/downloads/mk61rp641} (2024/31/01).

\bibitem{Scikit-learn}
F.~Pedregosa, G.~Varoquaux, A.~Gramfort, V.~Michel, B.~Thirion, O.~Grisel, M.~Blondel, P.~Prettenhofer, R.~Weiss, V.~Dubourg, J.~Vanderplas, A.~Passos, D.~Cournapeau, M.~Brucher, M.~Perrot, and E.~Duchesnay, ``Scikit-learn: Machine learning in {P}ython,'' \emph{Journal of Machine Learning Research}, vol.~12, pp. 2825--2830, 2011.

\bibitem{Frank:2005aa}
\BIBentryALTinterwordspacing
E.~Frank, M.~A. Hall, G.~Holmes, R.~Kirkby, B.~Pfahringer, and I.~H. Witten, \emph{{Weka: A machine learning workbench for data mining.}}\hskip 1em plus 0.5em minus 0.4em\relax Berlin: Springer, 2005, pp. 1305--1314. [Online]. Available: \url{http://researchcommons.waikato.ac.nz/handle/10289/1497}
\BIBentrySTDinterwordspacing

\bibitem{Chandola_2009}
\BIBentryALTinterwordspacing
V.~Chandola, A.~Banerjee, and V.~Kumar, ``Anomaly detection: A survey,'' \emph{ACM Comput. Surv.}, vol.~41, no.~3, jul 2009. [Online]. Available: \url{https://doi.org/10.1145/1541880.1541882}
\BIBentrySTDinterwordspacing

\bibitem{Sayadi_2017}
H.~Sayadi, N.~Patel, A.~Sasan, and H.~Homayoun, ``Machine learning-based approaches for energy-efficiency prediction and scheduling in composite cores architectures,'' in \emph{2017 IEEE International Conference on Computer Design (ICCD)}, 2017, pp. 129--136.

\bibitem{VirusTotal}
Chronicle, ``Virustotal,'' Available at {https:// www.virustotal.com/gui/home/upload} (2023/06/29).

\bibitem{Qiao:2016aa}
Y.~Qiao, X.~Yun, and Y.~Zhang, ``How to automatically identify the homology of different malware,'' in \emph{2016 IEEE Trustcom/BigDataSE/ISPA}, 2016, pp. 929--936.

\bibitem{Hu_2015}
H.~Hu, Z.~L. Chua, S.~Adrian, P.~Saxena, and Z.~Liang, ``Automatic generation of data-oriented exploits,'' in \emph{Proceedings of the 24th USENIX Conference on Security Symposium}, ser. SEC'15.\hskip 1em plus 0.5em minus 0.4em\relax USA: USENIX Association, 2015, pp. 177--192.

\bibitem{Konstantinou:2022aa}
C.~Konstantinou, X.~Wang, P.~Krishnamurthy, F.~Khorrami, M.~Maniatakos, and R.~Karri, ``Hpc-based malware detectors actually work: Transition to practice after a decade of research,'' \emph{IEEE Design \& Test}, vol.~39, no.~4, pp. 23--32, 2022.

\bibitem{Guthaus_2001}
M.~Guthaus, J.~Ringenberg, D.~Ernst, T.~Austin, T.~Mudge, and R.~Brown, ``Mibench: A free, commercially representative embedded benchmark suite,'' in \emph{Proceedings of the Fourth Annual IEEE International Workshop on Workload Characterization. WWC-4 (Cat. No.01EX538)}, 2001, pp. 3--14.

\bibitem{Dietterich_2000}
T.~G. Dietterich, ``Ensemble methods in machine learning,'' in \emph{Multiple Classifier Systems}.\hskip 1em plus 0.5em minus 0.4em\relax Berlin, Heidelberg: Springer Berlin Heidelberg, 2000, pp. 1--15.

\bibitem{Freund_Schapire_1997}
\BIBentryALTinterwordspacing
Y.~Freund and R.~E. Schapire, ``A decision-theoretic generalization of on-line learning and an application to boosting,'' \emph{Journal of Computer and System Sciences}, vol.~55, no.~1, pp. 119--139, 1997. [Online]. Available: \url{https://www.sciencedirect.com/science/article/pii/S002200009791504X}
\BIBentrySTDinterwordspacing

\bibitem{Sayadi_2020}
\BIBentryALTinterwordspacing
H.~Sayadi, Y.~Gao, H.~Mohammadi~Makrani, T.~Mohsenin, A.~Sasan, S.~Rafatirad, J.~Lin, and H.~Homayoun, ``Stealthminer: Specialized time series machine learning for run-time stealthy malware detection based on microarchitectural features,'' in \emph{Proceedings of the 2020 on Great Lakes Symposium on VLSI}, ser. GLSVLSI '20.\hskip 1em plus 0.5em minus 0.4em\relax New York, NY, USA: Association for Computing Machinery, 2020, pp. 175--180. [Online]. Available: \url{https://doi.org/10.1145/3386263.3407585}
\BIBentrySTDinterwordspacing

\end{thebibliography}

\begin{IEEEbiography}[{\includegraphics[width=1in,height=1.25in,clip,keepaspectratio]{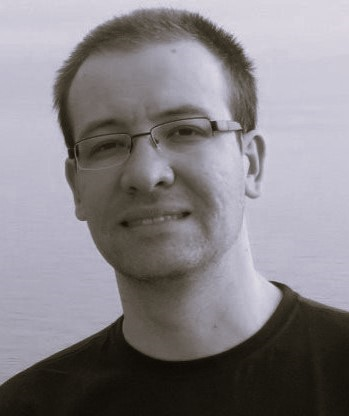}}]{Cristiano Pegoraro Chenet}
is a Ph.D. candidate at the Department of Computer and Control Engineering of Politecnico di Torino. His current research focuses on cybersecurity. He is a student member of IEEE.
\end{IEEEbiography}

\begin{IEEEbiography}[{\includegraphics[width=1in,height=1.25in,clip,keepaspectratio]{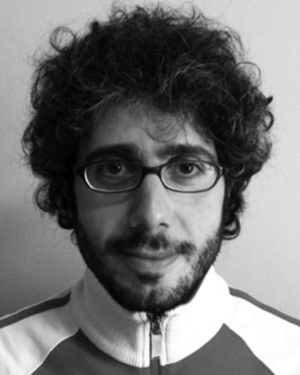}}]{Alessandro Savino}
is an associate professor at the Department of Control and Computer Engineering, Politecnico di Torino, 10129 Turin, Italy. His research interests include approximate computing, reliability analysis, safety-critical systems, software-based self-test, operating systems, imaging algorithms, machine learning, and audio manipulation. Savino received his Ph.D. from the Politecnico di Turin. He is a senior member of IEEE.
\end{IEEEbiography}

\begin{IEEEbiography}[{\includegraphics[width=1in,height=1.25in,clip,keepaspectratio]{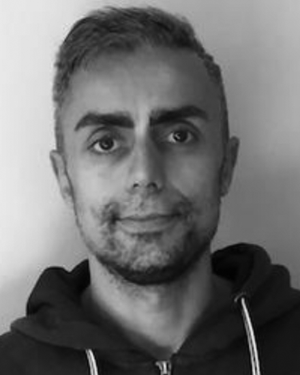}}]{Stefano di Carlo}
 received the MS equivalent and PhD degrees in computer engineering and information technology from the Politecnico di Torino in Italy in 1999 and 2003, respectively. Since 2021, he has been a full professor at the Department of Control and Computer Engineering, Politecnico di Torino, Italy. His research interests encompass a diverse range, including reliability analysis, FPGA design, memory testing, NVM memory reliability with ECC, design for testability, built-in self-test, fault simulation, and automatic test generation. With over 200 peer-reviewed publications in esteemed IEEE/ACM transactions, journals, and conference proceedings, he also contributes to the Editorial Board of top-tier journals. His involvement extends to serving on various Organizing and Program committees for major IEEE and ACM conferences and Symposia. Notably, he is recognized as a Golden Core member of the IEEE Computer Society and has received both Outstanding and Meritorious Awards for his volunteer efforts within the society. He holds the distinguished status of senior member within IEEE.
\end{IEEEbiography}

\EOD

\end{document}